\documentclass[fleqn,usenatbib]{mnras}

\usepackage{newtxtext,newtxmath}

\usepackage[T1]{fontenc}

\DeclareRobustCommand{\VAN}[3]{#2}
\let\VANthebibliography\thebibliography
\def\thebibliography{\DeclareRobustCommand{\VAN}[3]{##3}\VANthebibliography}

\usepackage{graphicx}	
\usepackage{amsmath}	
\usepackage{pifont}
\usepackage{mathtools}
\usepackage{placeins}
\usepackage{float}
\usepackage{multirow}
\usepackage[table]{xcolor}
\usepackage{bm}
    
\usepackage{nccmath}

\usepackage{cleveref}
\crefname{figure}{Fig.}{Figs.}
\crefname{table}{Table}{Tables}

\title[Moving-mesh non-ideal MHD protostar formation]{Moving-mesh non-ideal magnetohydrodynamical simulations of the collapse of cloud cores to protostars}

\author[Mayer et al.]{%
Alexander C. Mayer$^{1,2}$\thanks{E-mail: amayer@mpa-garching.mpg.de}, Oliver Zier$^{1,3, 4}$, Thorsten Naab$^{1}$, Rüdiger Pakmor$^{1}$, Paola Caselli$^{5}$, Alexei V. Ivlev$^{5}$, 
\newauthor Volker Springel$^{1}$ \& Stefanie Walch$^{6}$\vspace*{0.1cm}\\%
$^{1}$Max-Planck-Institut für Astrophysik, Karl-Schwarzschild-Str. 1, 85741 Garching, Germany\\%
$^{2}$Excellence Cluster ORIGINS, Boltzmannstraße 2, 85748 Garching, Germany\\%
$^{3}$Center for Astrophysics | Harvard \& Smithsonian, 60 Garden St, Cambridge, MA 02138, USA\\%
$^{4}$Department of Physics, Kavli Institute for Astrophysics and Space Research, Massachusetts Institute of Technology, Cambridge, MA 02139, USA\\%
$^{5}$Max-Planck-Institut f\"ur Extraterrestrische Physik, Giessenbachstr. 1, 85748 Garching, Germany\\%
$^{6}$ Universit\"at zu K\"oln, I. Physikalisches Institut, Z\"ulpicher Str. 77, 50937 K\"oln, Germany
}

\date{Accepted XXX. Received YYY; in original form ZZZ}

\pubyear{2024}

\begin{document}
 \label{firstpage}
\pagerange{\pageref{firstpage}--\pageref{lastpage}}
\maketitle

\begin{abstract}
Magnetic fields have been shown both observationally and through theoretical work to be an important factor in the formation of protostars and their accretion disks. Accurate modelling of the evolution of the magnetic field in low-ionization molecular cloud cores requires the inclusion of non-ideal magnetohydrodynamics (MHD) processes, specifically Ohmic and ambipolar diffusion and the Hall effect. These have a profound influence on the efficiency of magnetic removal of angular momentum from protostellar disks and simulations that include them can avoid the `magnetic-braking catastrophe' in which disks are not able to form. However, the impact of the Hall effect, in particular, is complex and remains poorly studied. In this work, we perform a large suite of simulations of the collapse of cloud cores to protostars with several non-ideal MHD chemistry models and initial core geometries using the moving-mesh code {\small AREPO}. We find that the efficiency of angular momentum removal is significantly reduced with respect to ideal MHD, in line with previous results. The Hall effect has a varied influence on the evolution of the disk which depends on the initial orientation of the magnetic field. This extends to the outflows seen in a subset of the models, where this effect can act to enhance or suppress them and open up new outflow channels. We conclude, in agreement with a subset of the previous literature, that the Hall effect is the dominant non-ideal MHD process in some collapse scenarios and thus should be included in simulations of protostellar disk formation.
\end{abstract}


\section{Introduction}
The evolution of young stars is intimately connected to the accretion disks surrounding them. They are thought to form along with the star, as the angular momentum of the collapsing molecular cloud core is distributed mostly outside of the central protostar in a rotating disk. However, high-angular resolution and high-sensitivity observations of protoplanetary disks in their early phases of evolution (Class 0/I) are still in their infancy \citep[e.g.][]{alves2019ybp,reynolds2021kinematic,maureira2022hotspots,tsukamoto2023ppvii,ohashi_tobin2023edisk}. There is also evidence of asymmetric accretion of material onto protoplanetary disks \citep[e.g.][]{pineda2020streamer,valdivia-mena2022streamer,hsieh2023streamer,flores2023streamer,podio2024streamer}, defying a simple picture of their formation. Therefore, their growth and accretion from the envelope are not yet understood in detail at this point. 

Theoretical work, pioneered by \cite{larson1969}, has however been successful in illuminating the main processes involved, suggesting that the formation of a stellar core roughly proceeds in four stages. During the initial collapse, the temperature is approximately constant, and gravity is only counteracted by the magnetic field, leading to the formation of a flattened `pseudo-disk' supported by magnetic forces, as established by \cite{Galli&Shu1993I,Galli&Shu1993II}. As gas accumulates, the densest region becomes optically thick and a pressure-supported region forms as the temperature rises, a structure known as the first (Larson) core \citep{hennebelle2008magnetic,xu2021formationI,xu2021formationII}. Further rise of the mass in this core brings the temperature up to the point of dissociation of H$_2$, which softens the equation of state and lets the collapse proceed further. Finally, the temperature evolution becomes adiabatic again and a new (smaller) pressure-supported region forms, called the second (Larson) core. This overall evolution has by now been thoroughly established by radiation (magneto-)hydrodynamical calculations \citep{whitehouse_bate2006collapse,bate2014collapse,tomida2013coreformation,tsukamoto2015radiative}, with other studies using barotropic equations of state based on the radiative calculations.

Magnetic fields are recognized to play a substantial role in this formation process. The strength of the magnetic field in the interstellar medium (ISM) down to molecular clouds and cloud cores has been measured, showing dynamically important levels of magnetization, where gas flows and the formation of structure are affected \citep[][and references therein]{Pattle2023} by it. Therefore, \textit{magnetic braking}, where the rotation of disks is slowed by the resistance to bending of magnetic field lines coupled to the gas, should be highly effective in the environment of protostellar accretion disks \citep[e.g.][]{allen2003braking,zhao2020review,yen_lee2024braking}. 

The degree of magnetization is usually characterized by the dimensionless mass-to-flux ratio:
\begin{equation}
\mu \equiv \frac{M / \Phi_{\rm{B}}}{(M / \Phi_{\rm{B}})_{\rm{crit}}},
\end{equation}
where $\Phi_{\rm{B}} \equiv \pi R^2 B$ is the magnetic flux (with $R$ referring to the radius of the core and $B$ the magnetic field strength) and 
\begin{equation}
(M / \Phi_{\rm{B}})_{\rm{crit}} \equiv \frac{c_1}{3 \pi} \sqrt{\frac{5}{G}},
\end{equation}
with $c_1 \approx 0.53$, is the critical value of the mass-to-flux ration in a spherically symmetric setup \citep{Mouschovias1976}, meaning that if $\mu > 1$, gravitational forces dominate and the core will collapse \citep{Shu1987}. The case $\mu = \infty$ corresponds to hydrodynamics.  However, the assumption of a complete coupling between gas and magnetic fields, which is the case in ideal magnetohydrodynamics (MHD) is not applicable, since the ionization degree is so low in these cloud cores that ionized particles can no longer fully drag the largely neutral gas along with it. This regime is called \textit{non-ideal} MHD. 

In a single-fluid simulation, the effects of decaying current due to finite conductivity of the gas (\textit{Ohmic resistivity}), the drift between charged and neutral species (\textit{ambipolar diffusion}) and the differences in coupling to the magnetic field of positive and negative charge carriers due to different inertia (\textit{Hall effect}) can be included via additional terms in the induction equation \citep[a detailed discussion is found in][]{Pandey2008}:
\begin{equation}
\label{eq:non_ideal_induction}
\frac{\partial {\bm B}_{\rm ni}}{\partial t} = - \nabla \times \{ \eta_{\rm OR} {\bm J} + \eta_{\rm HE}({\bm J} \times {\bm b}) - \eta_{\rm AD} [({\bm J} \times {\bm b}) \times {\bm b}]\},
\end{equation}
where ${\bm b} = {{\bm B}}/{\vert {\bm B} \vert}$ is the direction vector of the magnetic field, ${\bm J} = \nabla \times {\bm B}$ and $\eta_{\rm OR}$, $\eta_{\rm HE}$ and $\eta_{\rm AD}$ are the coefficient for Ohmic resistivity, the Hall effect and ambipolar diffusion, respectively. 

These three effects are strongest in different regions and phases of the process of star and disk formation, but there are also large regions of overlap where more than one effect is important \citep{wurster2021doweneed}. Generally speaking, ambipolar diffusion, which dominates in the early stages of collapse and still is the strongest effect far from the protostar in later phases, works against the bending of magnetic field lines and thereby reduces the buildup of the magnetic field in and near the protostar with respect to ideal MHD. Ohmic resistivity instead is only dynamically important at very high densities, and thereby only after the formation of the first Larson core. Both of these diffusive effects promote disk formation by reducing magnetic braking, while the influence of the Hall effect is more nuanced, as it can hinder or enhance the removal of angular momentum, depending on the local chemistry and geometry. With the inclusion of non-ideal MHD, the `magnetic braking catastrophe' can be avoided \citep[e.g.][]{Zhao2016,vaytet2018protostellar,wurster2021nonidealimpactsingle}.

Because the Hall coefficient $\eta_{\rm HE}$ is negative throughout most of the early evolution of the collapse of a rotating cloud core to a protostar and disk, the azimuthal Hall drift of magnetic field lines (which can be expressed as a drift velocity \textbf{v}$_{\rm HE}\propto - \eta_{\rm HE} {\bm J}$ in the non-ideal induction equation \ref{eq:non_ideal_induction}) is in the same direction as the azimuthal current, where the direction of the latter is set via the mostly hourglass-shaped magnetic field in the central pseudo-disk \citep{zhao2020hall}. Since the current is flowing against the rotation direction if the magnetic field is (close to) anti-aligned, in this case the efficiency of magnetic braking is reduced with respect to an identical situation without the Hall effect. In fact, the drift may be strong enough that the usual phenomenon of the extraction of angular momentum from the disk-plane is reversed, and the regions above and below the disk start to counter-rotate with respect to the envelope and initial conditions. 

Numerical modelling of the collapse of cloud cores to stellar core densities is inherently challenging for simulation codes, as a large range of scales has to be considered -- 10$^3$ to 10$^{-3}$ AU at the very least. In smoothed-particle-hydrodynamics (SPH) codes, the particles mass has to be set such that the Jeans length is sufficiently resolved at the highest density considered, which in general leads to a relatively low number of particles within the central region as a fraction of the total particle number. Adaptive-mesh-refinement (AMR) or nested grid codes can more effectively concentrate resolution elements within the core region, but suffer from the requirement of a very deep level hierarchy. For all codes, the inclusion of non-ideal MHD requires an extra criterion for the timesteps in the simulation, further increasing computational cost as more steps are required compared to the already expensive ideal MHD.

At this point, the ability to treat non-ideal MHD has been added to a large range of astrophysical codes using different underlying methods, from SPH codes \citep{wurster2014ambipolar,wurster2016cannonidealmhd,tsukamoto2017hall}, meshless-finite-mass/volume \citep{hopkins2017diffusion} to static, nested grid and AMR codes \citep{lesur2014thanatology,masson2012ambipolar,marchand2018hall,bai2011ambipolar,bai2014hall,stone2020athena++,li2011non}. Recently, it has been added to the moving-mesh code {\small AREPO} \citep{AREPODiffusion,AREPOHall} used in this work.

The coefficients for the magnetic diffusivity are strongly affected by the local ionization rate. However, there is so far only a limited number of studies which consider a number of different ionization rates and analyze the effect of this parameter on the properties of the resulting protostellar disks. This is despite the fact that large variations in cosmic-ray density are expected depending on position in the galaxy \citep[see e.g.][and references therein]{hanasz2021cosmicrays} and even within one star formation region \citep{Favre2018,Axen2024,pineda2024ngc1333}. \cite{wurster2018ionizationrates} tested a number of different ionization rates (15 values in the range $\zeta_i = 10^{-10}-10^{-30}$ s$^{-1}$), but used a relatively low resolution and only focused on the early phase of the collapse. \cite{Zhao2016} and \cite{zhao2020hall} considered a substantial number of different chemical models, grain size distributions and ionization rates in 2D simulations with the Hall effect. As they show, the grain size distribution has a significant effect on the diffusivity coefficients and thereby the evolution of their disks. \cite{tsukamoto2020dustmodel} similarly find the dust size distribution to be an important ingredient in setting disk size. 

In addition to non-ideal MHD, misalignment of initial magnetic field and rotation axis of the cloud core has also been established as a mechanism to reduce magnetic braking and thereby facilitate the formation of disks \citep{hennebelle2009misalignment,joos2012misalignment}, as braking is significantly more effective if rotation is perpendicular to the magnetic field. Misaligned configurations have also been studied with the addition of non-ideal MHD effects \citep{masson2012ambipolar,tsukamoto2018misalignment,machida2020misalignment,Hirano2020misalignment}, and the impact on outflows has been an early focus \citep{matsumoto2014direction}. \cite{machida2020misalignment} find that while the direction of jets is strongly affected by the initial misalignment, the total mass and momentum in the outflow are not changed much, unless there is a full $90^\circ$ offset. As pointed out by \cite{tsukamoto2018misalignment}, the results of previous studies occasionally do not seem in agreement, and the impact of misalignment differs from ideal to non-ideal MHD. \cite{tsukamoto2017hall} performed simulations with different initial orientations of magnetic field and rotation axis until second core formation, both with and without the Hall effect, showing the dynamical importance of the Hall effect in a variety of collapse scenarios. \cite{wurster2021nonidealimpactsingle} compared simulations including only a subset of the non-ideal MHD effects that were all run until stellar core formation, in particular stressing the significance of the Hall effect, while \cite{vaytet2018protostellar} performed a comparison of ideal MHD and non-ideal MHD with only Ohmic and ambipolar diffusion at very high resolution and discuss the differences in the resulting stellar cores in detail. The simulations of \cite{machida2011circumstellar} also employ a very fine nested grid and investigate both non-magnetized models as well as non-ideal MHD in the form of Ohmic diffusion. \cite{ahmad2023protostar} have simulated the formation of the stellar core and followed its evolution over a comparatively long time in a radiation-hydrodynamics calculation. Recently, the formation of protostellar disks in low-metallicity environments has also been investigated \citep{sadanari2023firststars}. 

In this work, we perform the first simulations of stellar core formation employing the moving-mesh code {\small AREPO} and aim to capture changes in disk size and the magnetic field morphology up to the formation of the second core as a function of the employed chemical model and initial geometry of the cloud core. We showcase the ability of the code to model the full collapse from the molecular cloud core to the second Larson core with the inclusion of non-ideal MHD effects. 

This paper is structured as follows. In Section~\ref{sec:methods}, we describe our numerical methodology while in Section~\ref{sec:setup} we detail the set-up of our initial conditions, the equation of state and chemistry. Therein, we also list our full suite of simulations, which are analysed in Section~\ref{sec:results}. In Section~\ref{sec:discussion} we discuss limitations of our model and compare our results to the literature. We summarize our results and present conclusion in Section~\ref{sec:summary}.

\section{Numerical methods} \label{sec:methods}
\subsection{Simulation code}
{\small AREPO} (\citealp{AREPO}, \citealp{pakmor2016convergene}; public release \citealp{AREPOpublic}) is a moving-mesh (magneto-)hydrodynamics code, which has been used in a large range of different astrophysical applications from cosmological simulations \citep[e.g.][]{Vogelsberger2014, mtng} to stellar mergers \citep{schneider2019mergers}. The original implementation of MHD in the code is described in \cite{MHDArepo}, although the divergence condition is now usually treated via the Powell-scheme as detailed in \cite{AREPOPowell}. While we refer to the above works for further detail, we briefly summarize the basic operation of the code here. {\small AREPO} uses a finite-volume approach in which the MHD-equations are solved on an unstructured grid obtained from the Voronoi tesselation of a number of mesh-generating points which move along with the local fluid flow. In this way, the code combines aspects of Eulerian and Lagrangian codes. In contrast to AMR, there is no level hierarchy, but only a single mesh, which can be helpful in problems with a large spatial hierarchy, such as the one at hand. Compared to SPH codes, {\small AREPO} offers a higher spatial resolution at a given mass resolution  and more flexible refinement operations. 

While {\small AREPO} has not been used for the particular problem of protostar formation up to the second core phase yet, its ideal MHD modules have been used extensively for simulations of galaxies and molecular clouds \citep[e.g.][]{pakmor2024auriga,tress2024cmz}. Recently, the code has been shown to accurately model the magnetorotational instability \citep{AREPOmri}, disk fragmentation \citep{arepoFragmentation} as well as cold rotating disks \citep{AREPOcolddisks}, all problems which are of obvious relevance to the formation of protostellar disks.

\subsection{Non-ideal MHD in {\small AREPO}}

A description of the numerical methods for the recently added non-ideal MHD effects can be found in \cite{AREPODiffusion} and \cite{AREPOHall} for the diffusive effects and the Hall effect, respectively. Briefly, the effects are added in an operator-split fashion, where the gradient of the magnetic field on the interface that is required for the flux-calculation is obtained via an overdetermined least-square-fit over the magnetic field values of all cells bordering the interface. We have made some slight modifications to the scheme; the reasons and concrete changes are detailed in Appendix \ref{app:modificationsNonIdealMHD}. 

We have also extended the scheme to use variable coefficients in every hydro-cell. The maximum timestep of a cell with radius $r = \left(\frac{3V}{4\pi}\right)^{1/3}$ is limited by the non-ideal MHD effects according to
\begin{equation}
\Delta t_{\rm OR,AD,HE} = C_{\rm OR,AD,HE}\frac{r^2}{\vert \eta_{\rm OR,AD,HE} \vert} ,\label{eq: timestep_constraint}
\end{equation}
with $C_{\rm OR,AD,HE} < 1$ being an analogue to the CFL number which we set to $C_{\rm OR,AD,HE} = 0.5$ throughout this work, as we have found this to be stable in all simulations if used together with an additional timestepping criterion:
\begin{equation}
    \tilde{\Delta t}_{\rm OR,AD,HE} = \frac{0.05}{3.0} \frac{\vert \bm{B} \vert}{\vert \nabla \bm{B} \vert} \frac{r}{\eta_{\rm OR,AD,HE}} \, ,
\end{equation}
which limits the change of the conserved magnetic field ($\bm{B}$ multiplied by volume) due to a non-ideal MHD effect to 5\% of the value before the calculation\footnote{The factor 3.0 derives from the ratio of volume and surface area of the volume-equivalent sphere.}. This additional criterion dominates over the standard one in regions of strong gradients and comparatively larger cells, and appears to help stabilize regions where non-ideal MHD is very strong. 

For the calculation of the flux over an interface, we utilize the harmonic mean of the two coefficients constituting the interface (as this ensures that equation \eqref{eq: timestep_constraint} is not violated in regions of strong differences in coefficients):
\begin{displaymath}
\eta_{\rm interface} = \frac{2\eta_1 \eta_2}{\eta_1 + \eta_2} \quad {\rm if} \quad \eta_1 \eta_2 > 0;
\end{displaymath}
\begin{displaymath}
\eta_{\rm interface} = 0 \quad {\rm else}.
\end{displaymath}
The latter case can only occur for the Hall effect and in this case the coefficients on opposite sides of the interface are both small anyway. 

Equation~\eqref{eq: timestep_constraint} easily dominates the other timestep constraints in the simulation (namely ideal MHD, with the usual CFL criterion including the fast magnetosonic speed, and gravity) and can lead to timesteps on the order of hours throughout a large timespan of the simulation. This only concerns a small fraction of cells, but substantially enlarges the timestep hierarchy, which leads to a significant runtime increase when non-ideal MHD is included. 

\section{General setup}
\label{sec:setup}
\subsection{Initial conditions}
The initial conditions are inspired by and similar to those used in \cite{wurster2018collapse}. In the initial state of the simulation, a core with a mass of $1\,{\rm M}_\odot$ and a radius of $r_0=3000$ AU (resulting in an initial uniform density of $\approx 5.25\times 10^{-18} {\rm g}\, {\rm cm}^{-3}$ and thereby a freefall time of $t_{\rm ff} \approx 29$ kyr) is placed within an ambient medium which is less dense by a factor of 30\footnote{We ensure initial pressure equilibrium between core and background despite the use of a barotropic equation of state by scaling up the pressure of the less dense background, representing a higher temperature in the surroundings of the core. This is done via a passive tracer field which gets advected with the infall, so a shock never develops at the boundary.}. The initially constant magnetic field is set such that it matches a desired mass-to-flux ratio within $r_0$. The chosen value is $\mu = 5$, which results in an initial field strength of $B_0 \approx 129 \mu {\rm G}$. The initial rotational velocity is $\Omega \approx  1.48\times 10^{-13} {\rm s}^{-1}$ (set to match a ratio of rotational to gravitational energy of $\beta_{\rm r} = 0.005$, which also matches \citealp{wurster2021nonidealimpactsingle} but is lower than used in e.g. \citealp{zhao2020hall}), where the core rotates as a solid body around the $z$-axis. The size of the box is chosen as $4 \times r_0$ to prevent effects of the boundary affecting the core. The boundary is periodic for hydro-calculations, but not for gravity. Cells are initialized on two closely-packed cubic latices, with a smaller spacing around the core (chosen such that the initial mass of the cells in the inner region is approximately equal to the target mass resolution) and a larger one outside.

\subsection{Resolution and refinement}
We rely on the standard method of refinement in {\small AREPO}, where a target mass for the gas is specified and cells are refined (split) or derefined (merged) if they deviate from this target by more than a factor of 2. Gas cells have a target mass of $3.33 \times 10^{-7}\, {\rm M}_\odot$ which is the same mass resolution as used in previous studies with Lagrangian codes \citep{wurster2018collapse,wurster2021nonidealimpactsingle,tsukamoto2017hall}. For a refinement criterion based on mass, resolving the second core requires having a number of resolution elements per minimum Jeans mass in the simulation, which for our equation of state is $\sim 10^{-4}$ M$_\odot$ at $\rho \sim 10^{-4}$ g cm$^{-3}$. The estimated number of required SPH particles per Jeans mass is $\sim$ 75 \citep{wurster2019catastrophe}, and since {\small AREPO} gives a higher spatial resolution at given target mass (note that there is no smoothing length), we easily resolve this minimum Jeans mass with $\sim 300$ cells. However, we do impose a minimum cell size of $r\approx 8 \times 10^{-5}$ AU \citep[similar as][]{vaytet2018protostellar}, but this only takes effect after the formation of the second core.

\subsection{Equation of state and non-ideal MHD chemistry}
We utilize a barotropic equation of state in this work, where the pressure (and thereby temperature) of a gas cell is completely determined as a function of density alone:
\begin{equation}
P = \rho c_{s,0}^2 \sqrt{1 + \frac{n} {n_1}}\left( 1 + \frac{n}{n_2}\right)^{-0.4}\left( 1 + \frac{n}{n_3}\right)^{0.37} ,
\label{eq: eos}
\end{equation}
where $n = \frac{\rho} {\mu m_{\rm p}}$ (where $m_{\rm p}$ is the proton mass) with $\mu \approx 2.381$ and $c_{s,0}=0.22\,{\rm km\, s}^{-1}$ to match the default parameters in the utilized chemical library (see below). This equation of state first has a transition from isothermal to adiabatic and then to $\gamma = 1.1$ to capture the dissociation of ${\rm H}_2$ and finally to adiabatic again for atomic hydrogen. We use $n_1 = 2.0 \times 10^{10}\, {\rm cm}^{-3}$, $n_2 = 2.5\times 10^{14}\,{\rm cm}^{-3}$ and $n_3 = 1.0 \times 10^{20}\,{\rm cm\textbf{}}^{-3}$, respectively, which matches one of the default barotropic equations of state in the {\small NICIL} library \citep{wurster2016nicil,wurster2021nicil2.0}. 

To obtain the non-ideal MHD coefficients $\eta_{\rm{OR},\rm{AD},\rm{HE}}$, we use pre-tabulated values from {\small NICIL}  between which we linearly interpolate given the density and magnetic field of a cell. We use the same equation of state in {\small NICIL} as we use for the gas, equation \eqref{eq: eos} above. We utilize different cosmic-ray ionization rates and grain size distributions throughout this work, as detailed in Section~\ref{subsection:performed simulations}. The interpolation from tables can be used with coefficients from any chemical library and can also easily be extended to non-barotropic equations of state, where one would additionally interpolate in the gas temperature. While the use of a table with a limited amount of values somewhat reduces the accuracy of the local resistivity calculation, the bigger limitation is likely the assumption of equilibrium chemistry, along with a fixed grain size distribution. 

The Hall effect fundamentally causes numerical issues in a wide range of methods used in astrophysical simulation codes \citep{bai2014hall,marchand2018hall,zhao2020hall}. \cite{AREPOHall} made use of the empirically determined threshold for stability
\begin{equation}
\eta_{\rm OR} > 0.2 \times \vert \eta_{\rm HE}\vert,
\label{stability_threshold}
\end{equation}
and introduced artificial Ohmic diffusion as needed to meet this condition. As one focus of this study is to investigate the implications of the Hall effect for disk formation, increasing Ohmic diffusion with respect to runs without the Hall effect makes it unclear which changes result from the presence of the Hall effect and which from higher diffusion. Therefore, we instead lower the Hall coefficient such that
\begin{equation}
\vert \eta_{\rm HE} \vert \leq 5\times {\rm max}(\eta_{\rm OR},\eta_{\rm AD})
\end{equation}
is always fulfilled. While this somewhat reduces the contribution of the Hall effect, we can still readily identify differences and make statements about where the impact would be even more pronounced without this limit. Note in particular that we keep the sign of the Hall coefficient consistent with the chemical model.

\subsection{Performed simulations}
\label{subsection:performed simulations}

\begin{table*}
\begin{tabular}{|c|c|c|c|c|c|c|c|c|}\cline{2-9}
\multicolumn{1}{c|}{} & \multicolumn{1}{c}{}& \multicolumn{7}{c|}{MHD model} \\\cline{2-9}
\multicolumn{1}{c|}{}& \multicolumn{1}{c|}{$\zeta_i$} & \multicolumn{1}{c|}{$\infty$} & \multicolumn{2}{c|}{10$^{-16}$ s$^{-1}$} & \multicolumn{2}{c|}{10$^{-17}$ s$^{-1}$} & \multicolumn{2}{c|}{10$^{-16}$ s$^{-1}$}\\\cline{2-9}
\multicolumn{1}{c|}{}& \multicolumn{1}{c|}{grains} & \multicolumn{1}{c|}{-} & \multicolumn{4}{c|}{MRN} & \multicolumn{2}{c|}{constant}\\\cline{2-9} 
\multicolumn{1}{c|}{}& {Hall?} & - & No & Yes & No & Yes & No & Yes \\\cline{1-9}
\multirow{5}{*}{Orientation} 
      & 00$^\circ$ & \cellcolor{blue!25} i00 & \cellcolor{blue!25}16mrn00N & \cellcolor{blue!25}16mrn00H & \cellcolor{blue!25}17mrn00N & \cellcolor{blue!25}17mrn00H & \cellcolor{blue!25}16con00N & \cellcolor{blue!25}16con00H \\\cline{2-9}
      & 45$^\circ$ & \cellcolor{blue!25}i45 & \cellcolor{blue!25}16mrn45N & \cellcolor{blue!25}16mrn45H & \cellcolor{blue!25}17mrn45N & \cellcolor{blue!25}17mrn45H & \cellcolor{blue!25}16con45N & \cellcolor{blue!25}16con45H \\\cline{2-9}
      & 90$^\circ$ & \cellcolor{blue!25}i90 & \cellcolor{blue!25}16mrn90N & \cellcolor{blue!25}16mrn90H & \cellcolor{blue!25}17mrn90N & \cellcolor{blue!25}17mrn90H & \cellcolor{blue!25}16con90N & \cellcolor{blue!25}16con90H \\\cline{2-9}
      & 135$^\circ$  & - & - & - & - & - & - & \cellcolor{blue!25}16con135H \\\cline{2-9}
      & 180$^\circ$  & - & - & - & - & - & - & \cellcolor{blue!25}16con180H \\\cline{1-9}
\end{tabular}
\caption{Summary of all 23 performed simulations. Here, $\zeta_i$ is the global cosmic-ray ionization rate ($\zeta_i =\infty$ coresponding to ideal MHD), and the orientation refers to the initial offset between magnetic field and rotation axis, where $00^\circ$ means anti-aligned. We perform simulations only including the diffusive non-ideal terms, and some which also add the Hall effect. All non-ideal MHD coefficients are calculated using the {\small NICIL} library \citep{wurster2016nicil} under the assumption of a truncated MRN grain size distribution or a constant grain size $a_0 = 0.1 \mu {\rm m}$. For more information see Section~\ref{subsection:performed simulations}.}
\label{Simulationtable}
\end{table*}

We investigate 3 different general models of the magnetic field evolution: ideal magnetohydrodynamics, only the two diffusive non-ideal effects, and all three effects. For the latter two, we consider three different chemical models (all coefficients are calculated via {\small NICIL} version 2.1.1). The first two use a global cosmic-ray ionization rate of $\zeta_i = 10^{-16} {\rm s}^{-1}$, one with a constant grain size of $a_0 = 0.1 \mu {\rm m}$ and one with grains divided into 5 size bins distributed as a truncated MRN \citep{mathis1977mrn} profile from $a_{\rm min} = 0.1 \mu {\rm m}$ to $a_{\rm max} = 0.25 \mu {\rm m}$ (which is the grain size profile used in \citealp{xu2021formationI} and \citealp{xu2021formationII}; it was tested first in \citealp{Zhao2016}). The final model uses $\zeta_i = 10^{-17} {\rm s}^{-1}$ with the truncated MRN profile from before, while the constant size distribution was too costly to run at the current time, as the non-ideal coefficients are significantly larger.

Our aim in designing this suite of simulations was to investigate a broad range of possible regimes for non-ideal MHD. Ideal MHD presents a simplified model, assuming (among others) an infinite ionization rate, while $\zeta_i = 10^{-16} {\rm s}^{-1}$ with MRN-distributed grains corresponds to a scenario where non-ideal MHD effects are noticeable, but not yet dominant. In the other two chemical models non-ideal MHD dominates, and the MRN model with $\zeta_i = 10^{-17} {\rm s}^{-1}$ gives significant ambipolar diffusion, while the model with constant grain sizes instead has higher Hall diffusivity coefficients \citep{wurster2016nicil,wurster2021nicil2.0,Zhao2016}. 

The influence of the alignment of initial rotation axis and magnetic field is investigated by performing simulations with different initial conditions in which the fields are anti-aligned (which will in the following be referred to as an angle of 0$^\circ$, and other degree values indicate deviation from this default), perpendicular ($90^\circ$), and in between these two cases equally in the $-z$ and $+x$ direction ($45^\circ$). In addition, for the runs with the most dominant Hall effect (the models with $\zeta_i = 10^{-16} {\rm s}^{-1}$ and a constant grain size distribution), we also consider a parallel configuration (labelled 180$^{\circ}$) and a model with $135^{\circ}$ (between aligned and the perpendicular configuration); bringing the total number of simulations to 23. We refer to the different models with the naming convention:
\begin{equation}
\begin{Bmatrix}
i\\
16\\
17
\end{Bmatrix}
\begin{Bmatrix}
{\rm mrn}\\
{\rm con}
\end{Bmatrix} 
\begin{Bmatrix}
00\\
45\\
90\\
135\\
180
\end{Bmatrix} 
\begin{Bmatrix}
{\rm N}\\
{\rm H}
\end{Bmatrix},
\end{equation}
for example ``17mrn45H'', where the content of the first brackets denotes the ionization rate (infinite/$10^{-16}{\rm s}^{-1}$/$10^{-17}{\rm s}^{-1}$), the second refers to MRN or constant grain size distribution (in the case of non-ideal MHD), the third gives the initial angle of the magnetic field with respect to the $-z$-axis, and the fourth indicates whether or not the Hall effect is included in this non-ideal MHD calculation. Whenever one or more of the 4 descriptors is absent when referring to models, we mean the group of models to which the other descriptors apply. All simulations are listed in Table \ref{Simulationtable}.

We run all simulations approximately until the formation of the second core at $\rho \approx 10^{-4}$ g cm$^{-3}$ (cf. again equation \eqref{eq: eos} above for the equation of state employed), but do not analyze the second core due to the limitations of our simplified temperature treatment. We do not use sink particles in this study. While they are useful in simulating the long-term behavior of the disk, they inevitably are a numerical approximation and their specific treatment may influence the outcome of the simulations in terms of disk size and mass \citep{machida2014sinks,xu2021formationI}. We leave their use in our simulations for later work, where we aim to employ a physically motivated sub-grid treatment for sink particles.

\section{Results}
\label{sec:results}
\begin{figure*}
    \centering
    \includegraphics[width=0.9\linewidth]{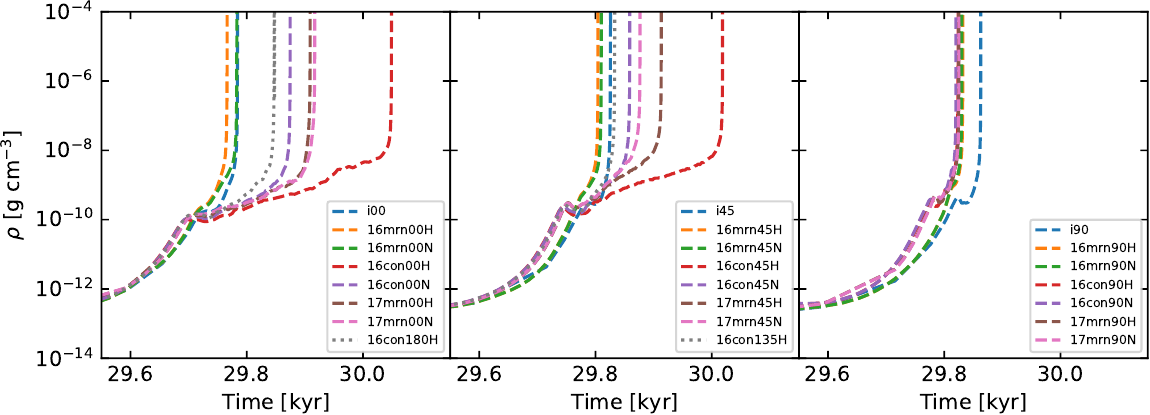}
    \caption{Evolution of the maximum density over time. The individual panels show the 3 different inclinations (left: 00$^{\circ}$, middle: 45$^{\circ}$, left: 90$^{\circ}$), where the 180$^{\circ}$ and 135$^{\circ}$ simulations are grouped with those with 0$^{\circ}$ and 45$^{\circ}$, respectively. The evolution starts to strongly deviate after a density of $10^{-10}$ g cm$^{-3}$ is reached. There is by far the least variation in the $90^\circ$ case.}
    \label{fig:dens_over_time}
\end{figure*}

\begin{figure*}
    \centering
    \includegraphics[width=0.9\linewidth]{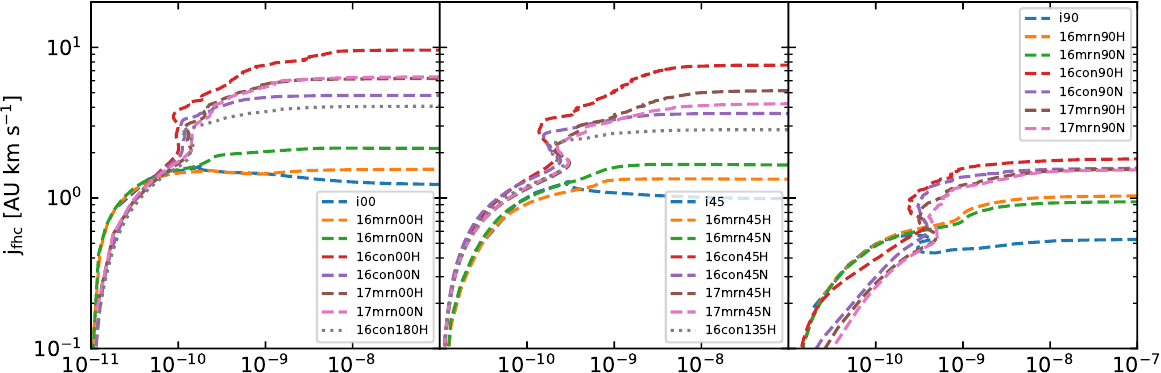}
    \includegraphics[width=0.9\linewidth]{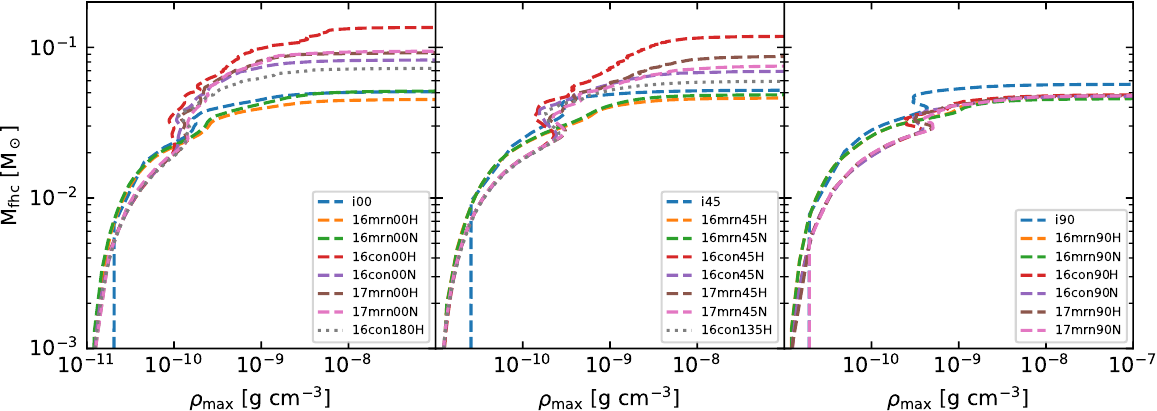}
    \caption{Evolution of the mean specific angular momentum (top) and mass (bottom) of the first hydrostatic core as a function of maximum density, which is used as a proxy for time. Stronger non-ideal MHD tends to increase the specific angular momentum and, with the exception of i90, the final mass of the first core.}
    \label{fig:fhc_over_time}
\end{figure*}

\begin{figure*}
    \centering
    \includegraphics[width=0.9\linewidth]{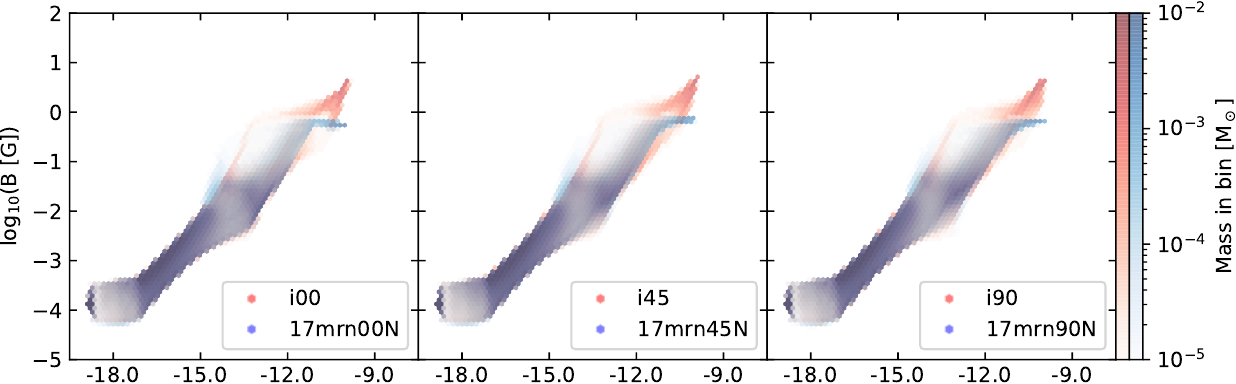}
    \includegraphics[width=0.9\linewidth]{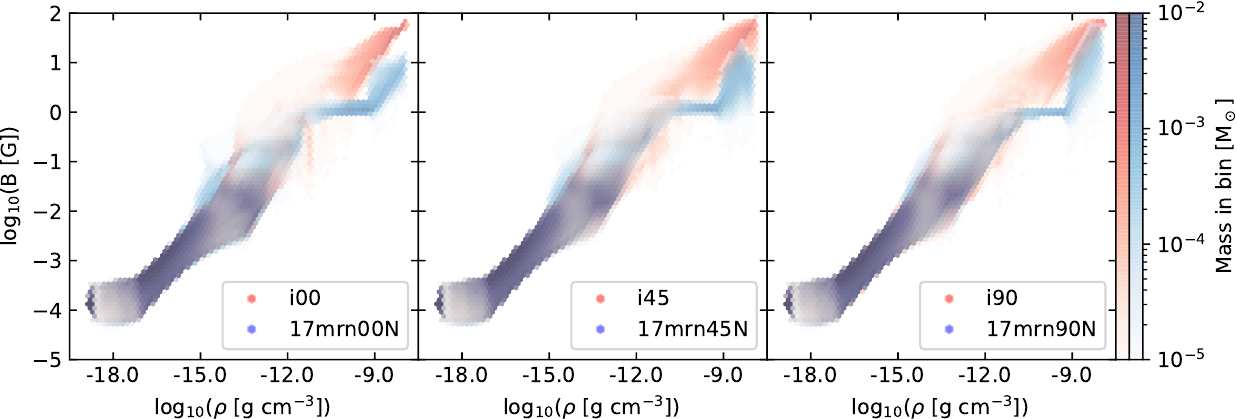}
    \caption{Histogram (colors indicate the total mass in each bin) of the $\rho$-$\vert{\bm B}\vert$ phase space; shown are ideal MHD and 17mrnN for the 3 different orientations (left to right: $00^\circ$, $45^\circ$, $90^\circ$)  at $\rho_{\rm max} = 10^{-10} \,{\rm g} \,{\rm cm}^{-3}$ (top) and $\rho_{\rm max} = 10^{-8} \,{\rm g} \,{\rm cm}^{-3}$ (bottom). The plateau in the non-ideal MHD models in contrast to ideal MHD can be clearly seen, but the magnetic field strength increases again at higher densities.}
    \label{fig:phasespace}
\end{figure*}

\begin{figure*}
    \centering
    \includegraphics[width=1\linewidth]{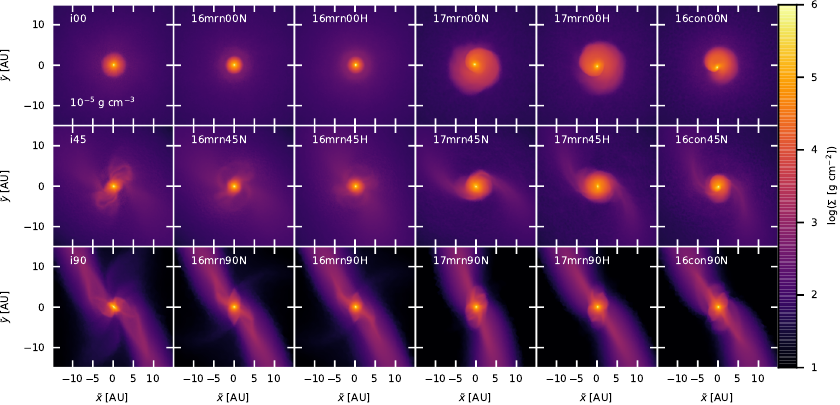}
    \includegraphics[width=1\linewidth]{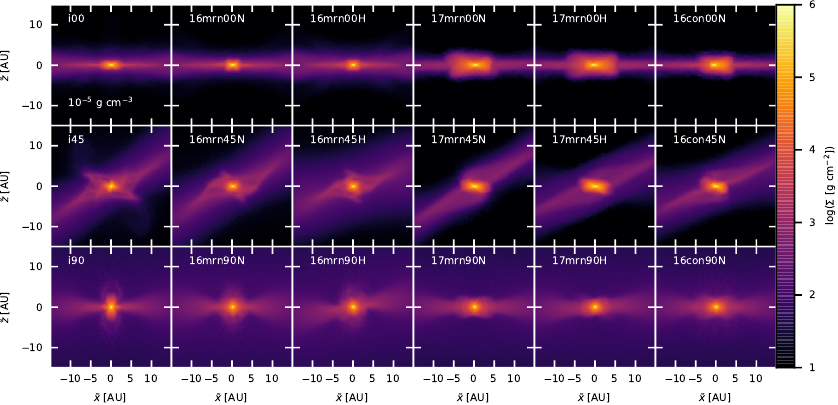}
    \caption{Top panels: Face-on projections (with respect to the angular momentum in the first hydrostatic core) close to second core formation for all but the 16conH models. Bottom panels: Edge-on projections. The rows contain simulations with the same initial magnetic field orientation, while the columns have the same chemical models (where the 3 columns on the left have much weaker non-ideal effects than the 3 on the right). Notice the tilt in the $90^\circ$ models when the Hall effect is included (column 3 and 5).}
    \label{fig:proj}
\end{figure*}

\begin{figure*}
    \centering
    \includegraphics[width=1\linewidth]{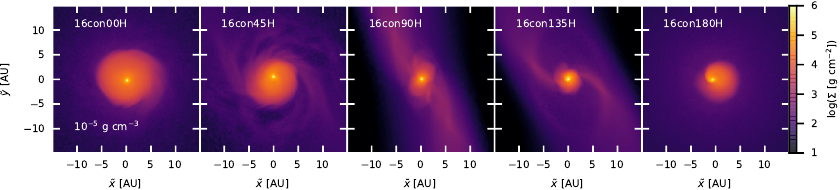}
    \includegraphics[width=1\linewidth]{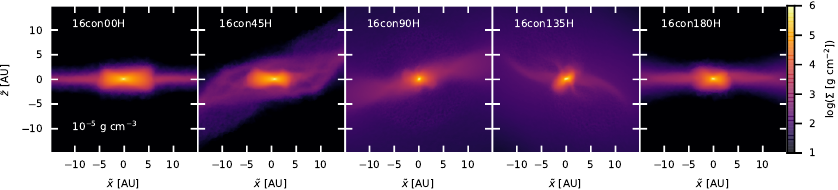}
    \caption{Face- and edge-on projections for the 16conH models. Here, the angle between initial magnetic field orientation and rotation axis becomes smaller from left to right. Differences between (largely) aligned and anti-aligned are seen both in the disk size and the structure of the pseudo-disk.}
    \label{fig:proj_hall}
\end{figure*}

\subsection{General evolution}
As a first indication of the evolution of the individual models, Figure~\ref{fig:dens_over_time} shows the maximum density among all resolution elements as a function of time. The different models collapse to the formation of the first core (defined as the point where a density of $10^{-11}$ g cm$^{-3}$ is reached) at approximately the same time ($t_{\rm fhc}\approx 29.7$ kyr), while the elapsed time for the formation of the second core (maximum density equal to $\approx 10^{-4}$ g cm$^{-3}$) is $\sim$ 29.75 to 30.05 kyr (deviation of $\approx$ 1\% of $t_{\rm ff}$). The evolution therefore appears to strongly deviate only after the formation of the first core, even though differences are visible early on, as we shall discuss. A second clear result is that the formation time of the second core is much more similar in the models with 90$^\circ$ inclination than for the others. 

Both of these conclusions appear to be related to the amount of angular momentum in the first core (defined as all material above $\rho_{\rm fhc} = 10^{-11}$ g cm$^{-3}$), which is displayed in the top part of Figure~\ref{fig:fhc_over_time} for all models: After the formation of the initially almost completely pressure-supported first core, the significant angular momentum in a subset of the models provides additional support against collapse, delaying the formation of the second core. In particular, the models with ideal MHD consistently show the lowest amount of specific angular momentum in the first core at the end of the simulation, and the angular momentum of the $90^\circ$-inclination runs are consistently much lower than for other alignments. The bimodality for disk evolution with the inclusion of the Hall effect is seen most clearly as the contrast in specific angular momentum between the 16con00H and 16con180H models on the one hand and 16con45H and 16con135H on the other, where the first simulation of each pair shows significantly higher values. Note that the corresponding models without the Hall effect -- 16con00N and 16con45N -- lie in between the two Hall models with opposite inclination. The 17mrnN models also show values in between, but higher than 16conN, implying that a dominant Hall effect can even be more effective at increasing the angular momentum of the first core than stronger (ambipolar) diffusion. 

A somewhat surprising observation in light of the the previously mentioned results is the faster collapse and lower specific angular momentum of 16mrn00H as compared to 16mrn00N, which also extends to 16mrn45N/H. This may be related to the fact that the Hall effect in this model is so weak that it really only has an effect in the surroundings of the disk instead of prominently in the disk itself -- and by reducing magnetic braking specifically in this region, it actually removes angular momentum from the disk plane. In 17mrn, we do not see the same scenario. This highlights the importance of the specifics of the chemical model for the evolution of the disk, and in particular of the modelling of the Hall effect. 

There appears to be a good correlation between the final mass in the first core (the evolution of this mass in shown in the bottom part of Figure \ref{fig:fhc_over_time}) and the time to collapse to a stellar core. This is not too surprising, as accretion is constantly happening and will give higher final mass if there is more time to accrete material -- although in some models there are outflows carrying away mass, which is quantified in subsection \ref{subsection:outflows}. On the other hand, as a comparison of the ideal MHD models in the $45^\circ$ and $90^\circ$ inclinations against the 16mrn models shows, the correlation between angular momentum in the first core and the final mass (or time) to collapse is less clear, as they have larger final masses but lower angular momentum. In this case, therefore, the rotational support is not the main contribution in delaying the collapse of the ideal MHD models. We further analyse the radial force distributions among multiple models in subsection \ref{subsection:radialforces}. First core masses at the time of stellar core formation are in the range of $5\times 10^{-2} - 2\times 10^{-1}\, {\rm M}_{\odot}$, and there is, as with the collapse times, much less variation between the different chemical models in the case of $90^{\circ}$, and their first core masses are at the low end of those seen for the other inclinations.

As can be seen in Figure~\ref{fig:phasespace}, the inclusion of non-ideal effects leads to a plateau in the magnetic field strength as a function of density in contrast to ideal MHD, which is at a strength $<\, 1{\rm G}$. This is a well-known result from earlier simulations \citep[e.g.][]{vaytet2018protostellar}. This plateau is then broken when the temperature gets sufficiently high for thermal ionization to be sufficient to cause a rapid decrease in the magnetic diffusivity coefficients, leading into a quasi-ideal-MHD regime in the densest region. In our simulations, the maximum magnetic field strength has already almost reached those of the simulations with ideal MHD at a density $\rho_{\rm max} = 10^{-8} \,{\rm g} \,{\rm cm}^{-3}$, and is basically identical at the formation of the second core. 

The large increase of the magnetic field after recoupling at high densities (Figure~\ref{fig:phasespace}) stands in contrast to earlier work where the magnetic field strength also increases above the plateau of the first core, but ends up being orders of magnitude weaker compared to ideal MHD \citep{wurster2018collapse,vaytet2018protostellar}. The reasons for this difference are as of now unclear, but could be related to our lack of a proper temperature treatment (as the equation of state becomes more and more crude at higher densities) which can affect when the magnetic field recouples and changes the plasma $\beta$, possibly affecting the magnetic field evolution. Another reason could be numerical effects originating in different resolutions or variations in the treatment of MHD by each code.

\subsection{Disk and pseudo-disk morphology}
\label{subsection:(pseudo-)disk}

It is well-established from previous simulations \citep[e.g.][]{xu2021formationI} that even in the absence of a rotationally supported disk a large flattened region, supported mainly by magnetic tension, forms perpendicular to the large-scale magnetic field, called a pseudo-disk. In the 00-models, the rotation is in the plane of this pseudo-disk, but in other orientations, the pseudo-disk is twisted by the rotation to some degree. To get an idea of the morphology of disk and pseudo-disk close to stellar core formation, Figures~\ref{fig:proj} and \ref{fig:proj_hall} show  projections of all models in two perpendicular directions: The new $z$-axis (denoted $\tilde{\bm z}$) used here is parallel to the angular momentum vector of the first core, while the $\tilde{\bm x}$-axis is defined as  \citep[similar to][]{tsukamoto2017hall}:
\begin{equation}
    \tilde{\bm x} = {\bm n}_{\rm pd} \times \tilde{\bm z},
\end{equation}
with ${\bm n}_{\rm pd}$ pointing along the direction of the eigenvector corresponding to the minimum eigenvalue of the inertial tensor of the pseudo-disk, which we define as all material with a density above $\rho_{\rm pd} = 10^{-14}$ g cm$^{-3}$. The plane for the edge-on projections in the 90$^\circ$ models is therefore usually close to that of the pseudo-disk.

These projections already give an impression about which models are similar in evolution, which will be repeatedly observed throughout the analysis. For the $00^\circ$ and $45^\circ$ inclinations (top and middle panels in each group of Figure~\ref{fig:proj}), there is a clear contrast between the simulations with ideal MHD and 16mrn, which are dominated by strong magnetic forces leading to outflows (clearly visible in the projections), and those which have comparatively much larger rotational support; namely the chemical models 17mrn and 16con. On the other hand, all of the previously mentioned simulations look quite distinct from those with $90^\circ$ initial misalignment (bottom panels of each group in Figure~\ref{fig:proj}). This appears to be somewhat in contrast with the findings of \cite{hennebelle2009misalignment}, who find large differences in ideal MHD simulations of disks misaligned by only $20^\circ$.

While the Hall effect is known to introduce a bimodality for disk evolution depending on the alignment of initial rotation and magnetic field axis by breaking the flip symmetry, this picture is further complicated when placing the initial axes at an angle from one another (see Figure~\ref{fig:proj_hall}). This can most clearly be seen in the lower group of panels of the Figure, which shows the disks of the Hall-dominated models edge-on. The $90^\circ$ simulation shows a distinctly warped pseudo-disk, and there are clear differences between the $45^\circ$ and $135^\circ$ cases as well as between the $00^\circ$ and $180^\circ$ inclinations, respectively. Even where non-ideal MHD effects are comparatively weak (specifically~16mrn), it can be seen that the Hall effect has exerted a torque in the innermost region of the $90^\circ$ simulation, which results in an offset in angle with its surroundings, in contrast to the model which includes only the diffusive effects.

In the edge-on projections, we can see that some of the simulations have become gravitationally unstable and have formed non-axisymmetric structures at this late point in evolution, consisting of a single spiral arm. Specifically, these are the simulations on the right side of Figure~\ref{fig:proj} (which have much stronger non-ideal MHD effects than those on the left) in the $00^\circ$ and $45^\circ$ configurations, and 16con00H, 16con45H and 16con180H for the Hall-dominated models. Gravitational instability will only develop if the disk is massive enough to locally collapse. This correlates, as discussed above in relation to the lower panels of Figure~\ref{fig:fhc_over_time}, with a slower collapse enabled by rotational support. The instability is also facilitated by relatively low magnetic pressure. As such, the spiral morphology occurs in our simulations where we expect it to, namely in the models with large angular momentum in the first core and strong diffusion. We shall further quantify the amount of rotational support of each simulation in subsection~\ref{Disk rotation velocities}. 

\begin{figure*}
    \centering
    \vspace{0 cm}
    \includegraphics[width=0.95\linewidth]{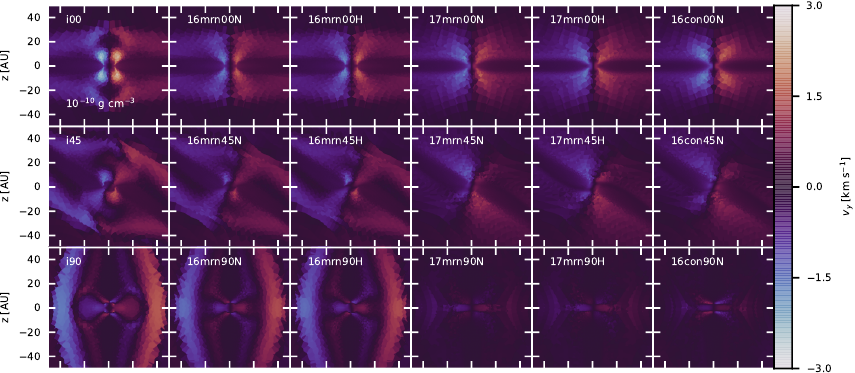}
    \includegraphics[width=0.95\linewidth]{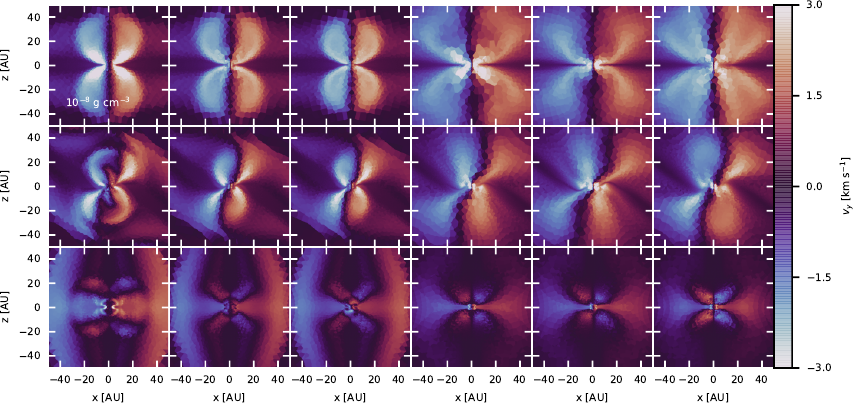}
    \caption{Slices of the line-of-sight velocity in the $xz$-place at the time the simulation reaches $\rho_{\rm max} = 10^{-10}, \, 10^{-8} \, {\rm g} \,{\rm cm}^{-3}$ (top, bottom) for the first time (all but the 16conH models). Note the presence of counter-rotating regions in the $90^\circ$ models, even without the inclusion of the Hall effect. When non-ideal MHD effects are dominant, rotation is fastest in the mid-plane in the later snapshots.}
    \label{fig:los_test}
\end{figure*}

\begin{figure*}
    \centering
    \includegraphics[width=1\linewidth]{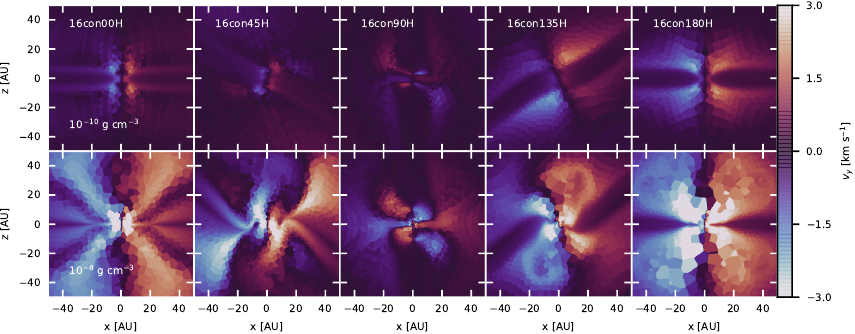}
    \caption{Vertical slices of the line-of-sight velocity for the Hall-dominated runs at  $\rho_{\rm max} = 10^{-10}, \, 10^{-8} \, {\rm g} \,{\rm cm}^{-3}$ (top, bottom). Note in particular the differences between the first and second column as opposed to the fourth and fifth column as well as the pronounced twist in the middle column.}
    \label{fig:los_hall}
\end{figure*}

\subsection{Rotational structure}
\label{subsection:Rotational structure}

In this subsection we discuss the general rotation structure of the models, while the occurrence of rotationally supported disks is the topic of the next subsection. Figures~\ref{fig:los_test} and \ref{fig:los_hall} show slices of the line-of-sight velocity in the $xz$-plane at different collapse times towards higher maximum densities (from top to bottom). Note that snapshots in one group of panels are not matched according to total elapsed time, but rather according to the same maximum density, thereby corresponding to the same `phase' of collapse. Already at a maximum  density of $10^{-10} \,{\rm g} \,{\rm cm}^{-3}$ (top panels), there are large differences between different chemical models. At this point, in the $00^\circ$ and $45^\circ$ orientations, the maximum line-of-sight velocity is not located in a central plane of rotation in the ideal MHD and 16mrn models, but is rather found above and below this plane. This is the typical result of strong magnetic braking and stands in contrast to what is seen in the 17mrnN/H and 16conN models. In the case of 16conH, the presence of this phenomenon strongly depends on the initial orientation. The rotation is fastest in the midplane for $00^\circ$ and $45^\circ$, and faster around it in the case of $135^\circ$ and $180^\circ$. Some regions are seen to be counter-rotating with respect to the original angular momentum and adjacent material. These occur where the magnetic tension force in the direction of counter-rotation is locally sufficiently strong to cause a reversal. Such regions appear in all of our $90^\circ$-models above and below the plane of the disk, while earlier work shows that a strong Hall effect can also produce them in the $00^\circ$ case \citep{wurster2021nonidealimpactsingle,tsukamoto2017hall}. 

In our results, the differences between $00^\circ$ and $180^\circ$ alignment, respectively $45^\circ$ and $135^\circ$ alignment, are still quite striking, and they are exclusively a consequence of the Hall effect (see Figure~\ref{fig:los_hall}). In 16con00, the surroundings are brought to almost no rotation (with their angular momentum instead being transported into the disk by magnetic forces), although the Hall effect is not strong enough to actually produce counter-rotation. Interestingly, however, 16con45H does show counter-rotation in these regions, which is not present in any of the other simulations with $45^\circ$ orientation. 

The Hall effect is also the cause for the twist in the rotation structure when the $90^\circ$ models with and without it are compared. This phenomenon is present for all chemical models, but is by far the most pronounced in 16con90H. The differences are smaller in the $45^\circ$ case, although they become larger as the simulation approaches stellar core formation for 17mrn45 and 16con45.

\begin{figure*}
    \centering
    \includegraphics[width=0.4\linewidth]{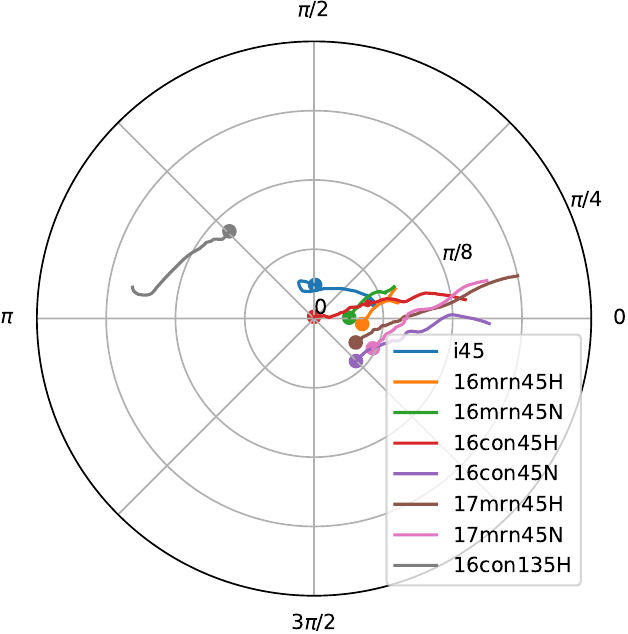}
    \includegraphics[width=0.4\linewidth]{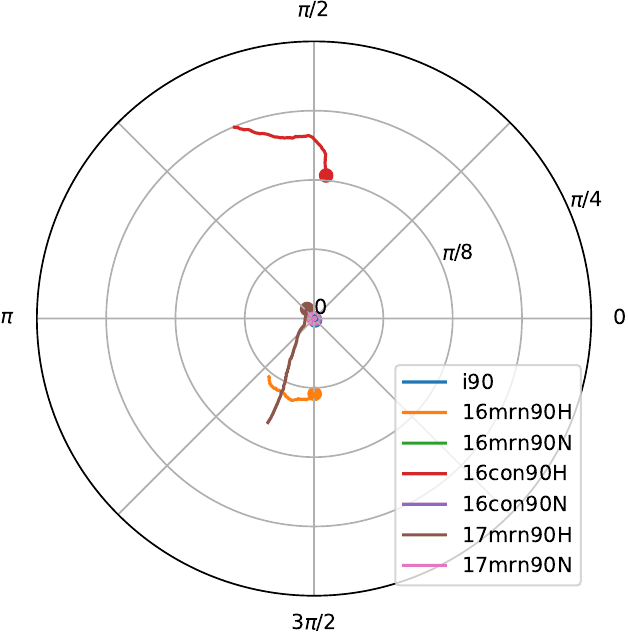}
    \caption{Time-evolution of the direction of the angular momentum vector in between maximum densities of $10^{-10} \,{\rm g} \,{\rm cm}^{-3}$ and $10^{-7} \,{\rm g} \,{\rm cm}^{-3}$ for inclinations of $45^\circ$ (left) and $90^\circ$ (right). The center corresponds to the positive $z$-direction, the distance from the center gives the polar angle and the position on the circle gives the azimuthal angle (with the positive $x$-direction located at 0). The dots mark the values at the highest density, which are essentially equal to those at the end of the simulation. There is more change in the case of $45^\circ$ inclination, and the angular momentum tends to evolve towards the rotation axis of the initial protostellar core.}
    \label{fig:3d_L}
\end{figure*}

Fast-rotating regions out of the plane of the disk often become part of the outflow through the course of the simulation (to be discussed in subsection~\ref{subsection:outflows}). They thereby effectively carry away angular momentum, and their suppression in some of the models allows for disks with faster rotation. In the models where initial rotation- and magnetic field axis are not (anti-)aligned, the direction of the mean angular momentum of the hydrostatic core changes substantially until the formation of the second core in some models, as seen in Figure~\ref{fig:3d_L}. This has previously been found and studied by \cite{matsumoto2014direction}. The direction appears to generally change more over the evolution of the disk in the $45^\circ$-models than in the $90^\circ$-models. Note, however, that Figure~\ref{fig:dens_over_time} indicates that this is likely related to there being less time between the formation of the first and second core. The angular momentum tends to gradually become more and more aligned with the $z$-axis, i.e.~the original angular momentum vector of the protostellar core. This might indicate that if we were to model the evolution of the disks for a longer period of time, they could potentially all end up close to the same axis of rotation, even though their early orientation differs substantially. While the rotational axis in 17mrnH does align itself with that of the core, 16mrn90H and 16con90H stay off. Interestingly, while all 45$^\circ$ inclination models start misaligned with the $z$-axis at $10^{-10} \,{\rm g} \,{\rm cm}^{-3}$, for 90$^\circ$ this is only the case if the Hall effect is included. This matches what is seen in Figures~\ref{fig:los_test} and \ref{fig:los_hall}, where the 90$^\circ$ models without the Hall effect show a complex structure of rotation and counter-rotation, but the central rotating region extends perpendicular to the $z$-axis without the Hall effect, which, if included, works to tilt it away from that plane.

\subsection{Disk rotation velocities}
\label{Disk rotation velocities}

\begin{figure*}
    \centering
    \includegraphics[width=1.0\linewidth]{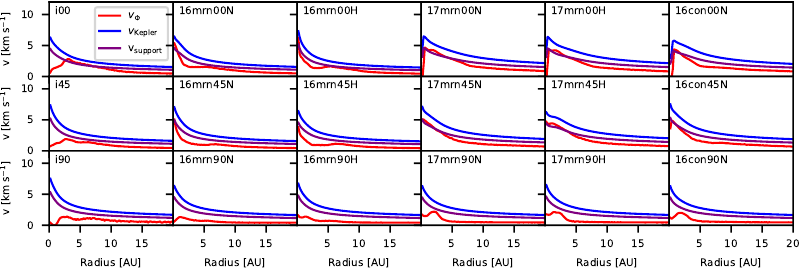}
    \includegraphics[width=1.0\linewidth]{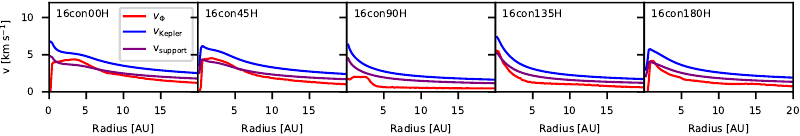}
    \caption{Radial profiles of the local azimuthal velocity. Some of the $90^\circ$ models may show a spiral-like morphology there (a twisted pseudo-disk), these clearly do not show the amount of rotational support that the other inclinations have. On the other hand, upper panels to the right have strong non-ideal MHD and, as a result, much rotational support.}
    \label{fig:radial_profile}
\end{figure*}

Only a subset of the simulations forms fast rotating disks, and whether they do depends both on the magnetic field model and the initial (mis-)alignment of magnetic field and rotation. The disks which do form also differ significantly in size. A qualitative picture of the disk sizes as well as their propensity to become gravitationally unstable can already be obtained from the face-on projections shown in Figures~\ref{fig:proj} and \ref{fig:proj_hall}.

To quantify the rotational support, in Figure~\ref{fig:radial_profile} we display the local azimuthal velocity (again in a frame that is turned to the angular momentum axis of the first core) as compared to the Keplerian velocity v$_{\rm Kepler}(r) = \sqrt{\frac{G M(<r)}{r}}$. The third velocity displayed in Figure~\ref{fig:radial_profile} is a metric for the rotational support in terms of the ratio of centrifugal and gravitational force in the plane of the disk \citep{tsukamoto2015q2b,tsukamoto2015q2a,wurster2021nonidealimpactsingle}:
\begin{equation}
q_2 (r) = \frac{\mathrm{v}_\Phi^2}{r} / \frac{G M(<r)}{r^2} ,
\end{equation}
and we set v$_{\rm Support}(r) = \frac{\mathrm{v}_{\rm Kepler}(r)}{\sqrt{2}}$, such that $q_2 (r)$ = 0.5 if the tangential velocity is equal to v$_{\rm Support}$ (for  $q_2 \geq 0.5$, the disk is primarily supported by the centrifugal force). Support is also provided by pressure forces, and another parameter ($q_1$) is often defined that includes both contributions (see references above), while $q_2$ specifically focuses on the rotational support. The $00^\circ$ models 17mrn00N/H and 16con00N/H show v$_\Phi \geq$ v$_{\rm Support}$ over a small radial range, mostly inside 5 AU, although some show fast rotation out to $\approx 10$ AU\footnote{The large velocities in i00 are likely because the azimuthal averaging picks up parts of the outflow, as the ``disk'' region is both thin and irregular in this simulation.}. Azimuthal velocities (compared to Keplerian) are generally lower for 45$^\circ$, but we again observe much more rotational support in the case of 17mrn45N/H and 16con45N/H. None of the simulations at $90^\circ$ magnetic field inclination have a rotationally supported disk, while 16con135H and 16con180H both do have substantial rotation, but less so than their counterparts 16con45H and 16con00H. These observations generally match the expectation from the projections in Figures \ref{fig:proj} and \ref{fig:proj_hall}, where the simulations with significant rotational support are those which have become gravitationally unstable close to stellar core formation. We do see a significant amount of rotational support in the inner 1 AU of the 16mrn00N/H models, which is likely an occurrence of the `second disk' that forms close to second core formation \citep{tomida2015seconddisk}.

Compared to, e.g., \cite{wurster2021nonidealimpactsingle} our disks appear to become gravitationally unstable more easily. This is likely related to our temperature treatment, as the barotropic equation of state tends to underestimate the disk temperature.

\subsection{Morphology and characteristic quantities of outflows}
\label{subsection:outflows}

 \begin{figure*}
    \centering
    \includegraphics[width=1\linewidth]{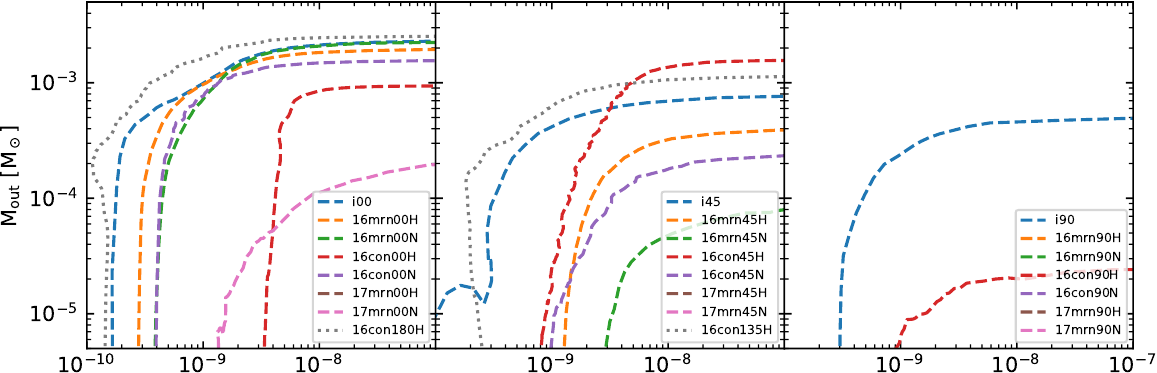}
    \includegraphics[width=1\linewidth]{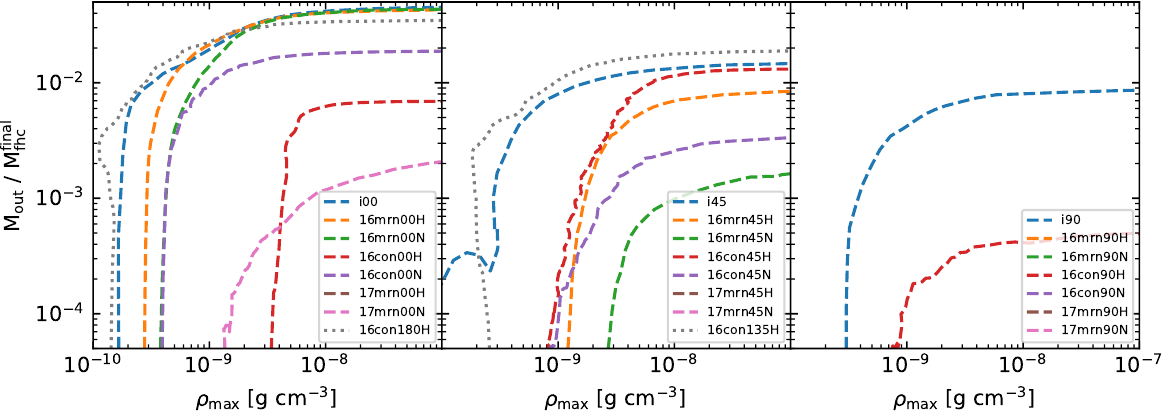}
    \caption{Outflowing mass of all models as a function of maximum density; the top panels show absolute values, while the bottom panels are scaled to the final mass of the first hydrostatic core in each simulation. Models with no visible lines do not produce outflows. All ideal MHD models show massive and early outflows, while non-ideal MHD effects tend to suppress them, just like an initial misalignment ($90^\circ$ much more strongly so than $45^\circ$).}
    \label{fig:outflow_quantities}
\end{figure*}

Some of the models are seen to produce magnetorotational outflows starting from the first core phase, which vary greatly in their strength and morphology. Due to our limitations in modelling the stellar core, we do not focus on stellar core outflows. To see which simulations produce outflows, we first need to have criteria to identify outflowing material. The main complication in this is that not all material with a negative radial velocity with respect to the core is actually outflowing. Instead, it can for example be part of a spiral arm travelling outwards. To exclude such material, we utilize both a density and velocity cut in restricting what is classified as part of the outflow: Material has to have a density less than $\rho_{\rm out} = 8\times 10^{-14}$ g cm$^{-3}$ (approximately the first transition point in the equation of state, where the pressure begins to rise) and a radial velocity larger than v$_{\rm out} = 0.1$ km s$^{-1}$. Unfortunately, it is essentially impossible to define simple criteria for outflows that include all clearly outflowing material while not including any of the disk. But we have seen that the above matches relatively well what is expected from a view of the radial velocity, while the disk is usually at higher density than what is allowed\footnote{Another frequently used criterion is a restriction to a range of angles above the disk plane, but this is problematic for arbitrary disk geometries.}. The evolution of the total outflowing mass at a given maximum density is shown in the top panels Figure~\ref{fig:outflow_quantities}.

In the 00$^\circ$-configuration, all models except 17mrn00H do produce outflows, while for 45$^\circ$ both 17mrn45H as well as 17mrn45N do not. The only model with a $90^\circ$ inclination and an outflow other than ideal MHD is 16con90H, but as compared to i90, the strength of the outflow is much reduced. The presence of an outflow in the case of a 90$^{\circ}$ misalignment in the ideal MHD model stands in contrast to earlier results from \cite{wurster2020turbI}, who used the same magnetization. The strongest outflows are seen in the 16con45H, 16con135 and 16con180H models. In the latter two, the Hall effect works to increase the efficiency of magnetic breaking by strengthening the azimuthal magnetic field in the case where initial magnetic field and rotation axis are (close to) anti-parallel. 

The differences between presence and absence of the Hall effect are also clearly seen in the 17mrn00N/H and 17mrn45N/H models, where the Hall effect reduces the amount of outflowing mass significantly. Consistent with the earlier somewhat puzzling findings of differences between 16mrn00N/H and 16mrn45N/H, the outflows in the models with the Hall effect are seen to carry more mass here. The phase of the collapse in which outflows start varies by more than an order of magnitude in terms of maximum density. Ideal MHD and 16con180/135H start the earliest, noticeable not long after $10^{-10}$ g cm$^{-3}$ is reached for the first time. Other simulations only show outflows after $10^{-9}$ g cm$^{-3}$ is reached, and there is some correlation between early outflows and the presence of large total masses, with a clear exception in the case of 16con45H. Recall that the same range of densities in Figure \ref{fig:outflow_quantities} can correspond to quite different lengths of time (Figure~\ref{fig:dens_over_time}), and generally, the simulations with more diffusion have both a longer first core lifetime and reduced outflows, and this leads to these two factors working against each other in determining the final outflow mass. Furthermore, since these outflows are produced by the twisting of the magnetic field, faster rotation (again linked to stronger diffusion) can promote outflows.

The mass of outflows of course depends in the first place on the amount of material that is available in the core. In the lower panels of Figure~\ref{fig:outflow_quantities} we therefore scale the total mass in the outflow displayed in the upper panels by the final mass in the first hydrostatic core (cf.~Figure~\ref{fig:fhc_over_time}). In this way, we can obtain a measure for how important the outflows are in regulating the mass budget of each system. Concerning the left panels, we observe that while 16con180H has the largest total outflowing mass, the relative mass is largest for i00 and 16mrnN/H, almost reaching 5\% of the mass of the first hydrostatic core at protostar formation. Similarly, 16con45H only dominates in total mass because it forms a more massive core, while i45 and 16con135H have larger scaled outflow masses. In the latter case, strengthened magnetic braking caused by the Hall effect makes the relative mass $\sim 1$\% higher than in the ideal case. However, for the 00$^\circ$ configuration the model 16con180H is not the simulation with the largest relative outflowing mass. For $90^\circ$, the relative masses, like the absolute masses before, are much lower than for the other orientations. In general, the $00^\circ$ inclination models follow the expectation in terms of relative outflow mass, where more diffusion reduces the mass, but the Hall effect can both decrease and increase it, and this in some cases becomes more important than the reduction through ambipolar diffusion. The $45^\circ$ models, however, do not show this trend, with the most surprising result being the higher mass in 16con45N compared to 16mrn45N, the latter of which is, in many ways, still much closer to ideal MHD than the former. The trends when the the Hall effect is included are also less clear, which indicates that not all conclusions from complete (anti-)alignment carry over to this inclined case.

\begin{figure*}
    \centering
    \vspace{0 cm}
    \includegraphics[width=0.95\linewidth]{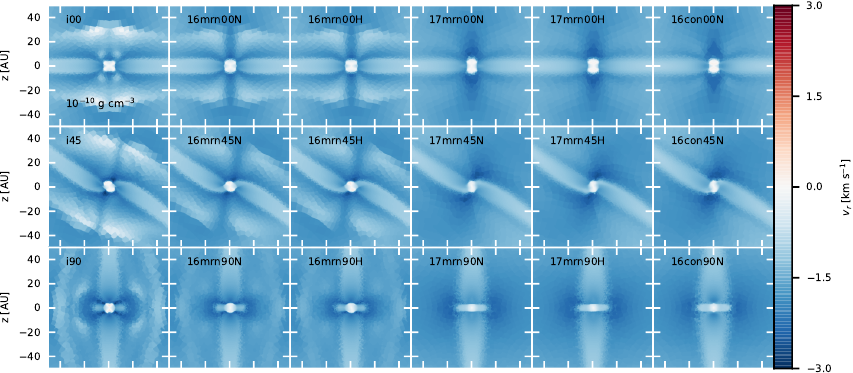}
    \includegraphics[width=0.95\linewidth]{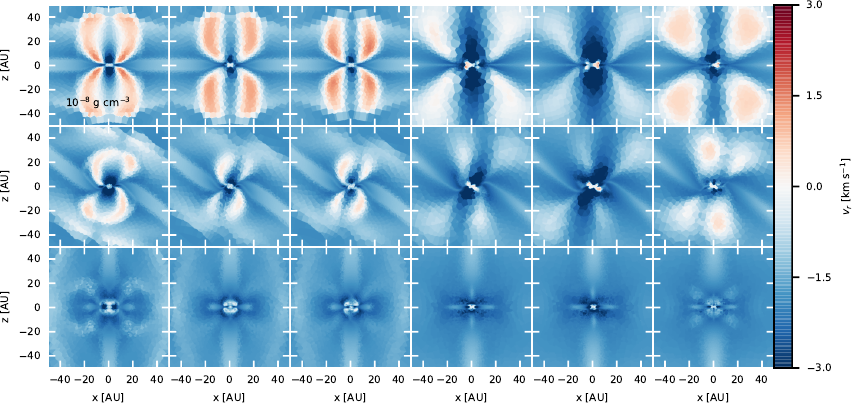}
    \caption{Slices of the radial velocity int the $xz$-plane at the time the simulation reaches $\rho_{\rm max} = 10^{-10}, \, 10^{-8} \, {\rm g} \,{\rm cm}^{-3}$ (top, bottom) for the first time. Outflows are the fastest and most collimated in the case of $00^\circ$ inclination, and when the simulation is closer to ideal MHD.}
    \label{fig:rad_test}
\end{figure*}

\begin{figure*}
    \centering
    \includegraphics[width=1\linewidth]{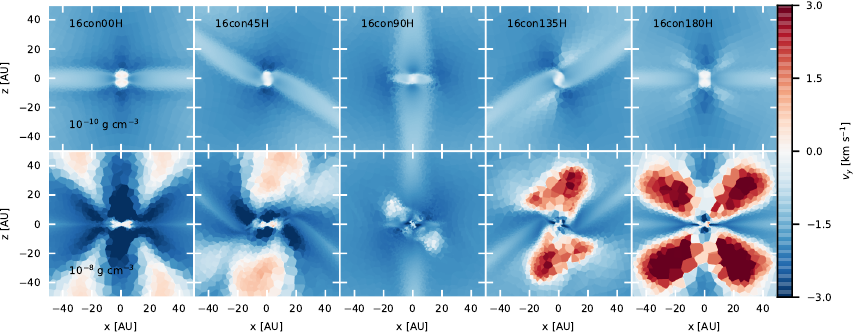}
    \caption{Slices of the radial velocity in the $xz$-plane for the Hall-dominated runs at  $\rho_{\rm max} = 10^{-10}, \, 10^{-8} \, {\rm g} \,{\rm cm}^{-3}$ (top, bottom). Note the large differences in outflow velocity in the first and second column as opposed to the fourth and fifth column in terms of velocity and extent of the outflow.}
    \label{fig:rad_hall}
\end{figure*}

Figures~\ref{fig:rad_test} and \ref{fig:rad_hall} show radial velocities in the $xz$-plane and thereby provide an overview of the outflow morphologies and velocities. There is significant evolution in the profiles shown. In the snapshots for 10$^{-10}$ g cm$^{-3}$, all simulations basically show the central first hydrostatic core with essentially zero radial velocity as the main feature, as well as the pseudo-disk as a region of reduced inflow velocity. As the density increases, outflow regions of strongly varying shapes and sizes start forming. Even early on, however, we can observe differences between chemical models with respect to where the inflow is fastest: In the $00^{\circ}$ models, the inflow velocities from above and below the core show large variations, with ideal MHD only having a fast inflow right next to the core, while the material further away cannot be accreted faster due to both its high angular velocity (cf.~Figure~\ref{fig:los_test}) and strong magnetic pressure (see Figure~\ref{fig:radial_forces} below). There is instead an X-shape (two truncated conical surfaces in 3D) around the core where the inflow is fastest. The inclusion of weak non-ideal MHD (16mrn) opens up a larger inflow channel that extends further out in the $z$-direction, while stronger diffusion leads to fast accretion from all directions except the pseudo-disk. 

Other orientations show a broadly similar picture in their corresponding directions perpendicular to the pseudo-disk. As outflows start to develop, usually from regions of fast rotation, we see that some chemical models produce relatively collimated outflows (ideal MHD and 16mrn) that mostly move in the $z$-direction in the $00^\circ$ configuration, while the other models -- if there are outflows -- show much broader outflow regions, moving in a more radial direction at a slower speed. Outflow speeds are lower in the $45^\circ$ models, and the axis of symmetry of the outflows is no longer perpendicular to the pseudo-disk, instead being somewhat more aligned with the $z$-axis. This is as expected from the inclination of the central rotating region, where, as discussed in subsection \ref{subsection:Rotational structure} (specifically relating to Figure~\ref{fig:3d_L}), the rotational axis is always inclined to a lesser degree with respect to the $z$-axis than the perpendicular vector of the pseudo-disk. For a $90^\circ$ inclination, we can again see that the formation of outflows is heavily suppressed. In the Hall dominated models, the higher relative mass of outflowing material of 16con180H with respect to 16con00H, and 16con135H to 16con45H, corresponds to a faster outward radial velocity. In the latter two, outflow regions are located closer to the hydrostatic core than in the former two.

\begin{figure*}
    \centering
    \includegraphics[width=1\linewidth]{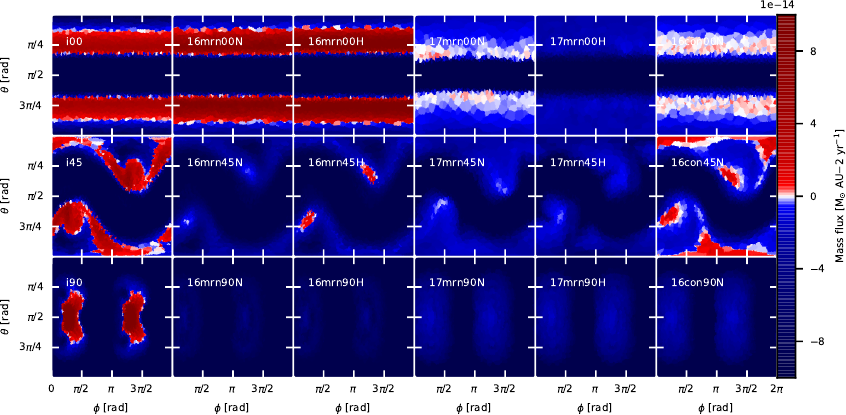}
    \caption{Rectilinear projection of mass flux at a distance of 25 AU at $\rho_{\rm max} = 10^{-8} \,{\rm g} \,{\rm cm}^{-3}$. Note the outflow in the i90 model, which is `missed' in the xz-slices. $\phi = 0,\pi$ corresponds to the plane shown in the $xz$-slices. Outflows tend to be highly localized, if present at all, when the initial orientation is not $00^\circ$.}
    \label{fig:contours}
\end{figure*}

\begin{figure*}
    \centering
    \includegraphics[width=1\linewidth]{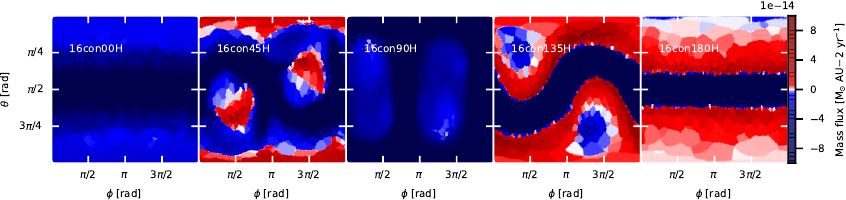}
    \caption{Rectilinear projection of mass flux at a distance of 25 AU at $\rho_{\rm max} = 10^{-8} \,{\rm g} \,{\rm cm}^{-3}$. The outflow in the 16con90H model is inclined compared to the one in the $i90$ simulation. 16con45H shows a relatively localized outflow with mostly inflow on this shell, while the reverse is true for 16con135H.}
    \label{fig:contours_hall}
\end{figure*}

To obtain a 3D view of the direction of outflows (since any 2D-slice can always `miss' them), Figures~\ref{fig:contours} and \ref{fig:contours_hall} show rectilinear projections of the mass flux through a shell located at a distance of 25 AU from the center. The lack of axisymmetry in the initial conditions extends to the structure of the outflows: In the case of model i90, it is principally aligned with the $y$-axis (perpendicular to both initial rotation axis and magnetic field), while the geometry is again more complex with the inclusion of the Hall effect.  We can now see the outflow in the i90 and 16con90H models, where the Hall effect has in the latter case acted to turn it out of the $xy$-plane. In the 16con45H model we see that at this distance from the center the outflow only happens through a relatively small channel, while in 16con135H material is inflowing from most directions outside the pseudo-disk. The aforementioned case of 16con90H is the most interesting, as none of the other non-ideal MHD models produces an outflow in this orientation, not even 16mrn90N/H, whose evolution we have seen to generally show similar behavior to ideal MHD. This implies that here the twist in the magnetic field (and, as a consequence, in the entire disk structure) caused by the Hall effect actually provides a new escape channel that is usually suppressed by such a strong misalignment of rotation axis and magnetic field. We show the corresponding three-dimensional morphology of the magnetic field in Section~\ref{subsection:magnetic_morphology}.

\subsection{Radial force contributions}
\label{subsection:radialforces}

\begin{figure*}
    \centering
    \includegraphics[width=1.0\linewidth]{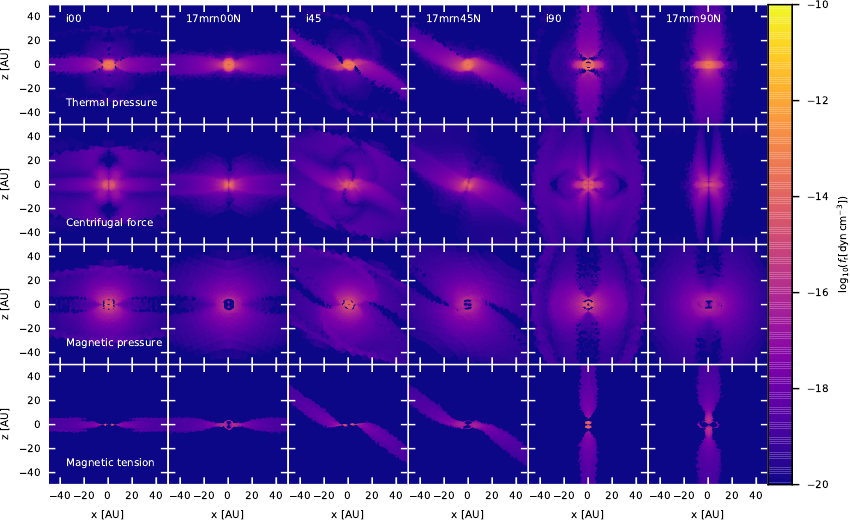}
    \caption{Radial force densities for ideal MHD and 17mrnN at $\rho_{\rm max} = 10^{-10}\,{\rm g\, cm}^{-3}$. Magnetic support is much stronger in the case of ideal MHD, in particular as the main force keeping material in the pseudo-disk from collapse. The $90^\circ$ simulations show relatively little rotational support.}
    \label{fig:radial_forces}
\end{figure*}

When discussing the general evolution of the models in comparison to each other, we have alluded to the fact that the differences in collapse times can be explained by the (lack of) support against gravitational collapse provided by centrifugal and magnetic forces. To quantify this support, we show in Figure \ref{fig:radial_forces} the radial components of the  thermal and centrifugal force density as well as those due to magnetic pressure and magnetic tension for the `extremes' of ideal MHD and an ionization rate of $10^{-17}$ s$^{-1}$ (17mrnN; without Hall effect in order to have the same symmetries) for the 3 considered initial orientations. In all cases, the pseudo-disk can be identified easily by the significant magnetic tension acting along it. The magnetic support is much stronger in the ideal MHD case, in particular still being strong in the hydrostatic core, for which magnetic tension support is essentially absent in the non-ideal MHD simulation. The centrifugal support traces the rotational structure as already seen in Figure~\ref{fig:los_test}. Since the centrifugal support appears to be subdominant in the case of $90^\circ$ orientation in the cnetral plane, the two kinds of magnetic support which are much stronger for ideal MHD appear to explain the delayed collapse of i90 with respect to the non-ideal MHD models.

\subsection{Analysis of magnetic field morphology}
\label{subsection:magnetic_morphology}
\begin{figure*}
    \centering
    \vspace{0 cm}
    \includegraphics[width=0.95\linewidth]{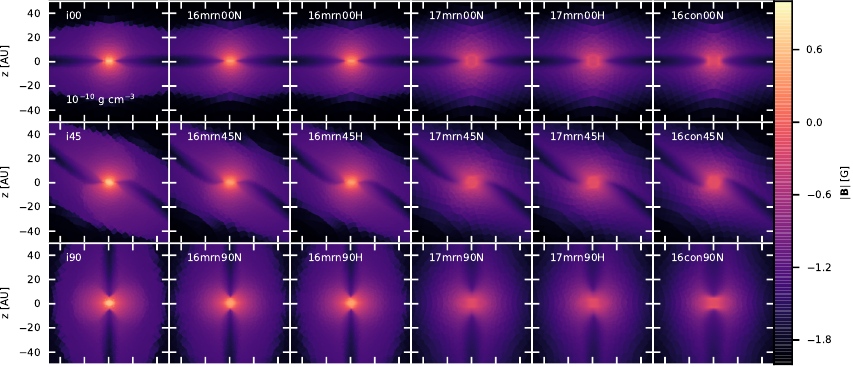}
    \includegraphics[width=0.95\linewidth]{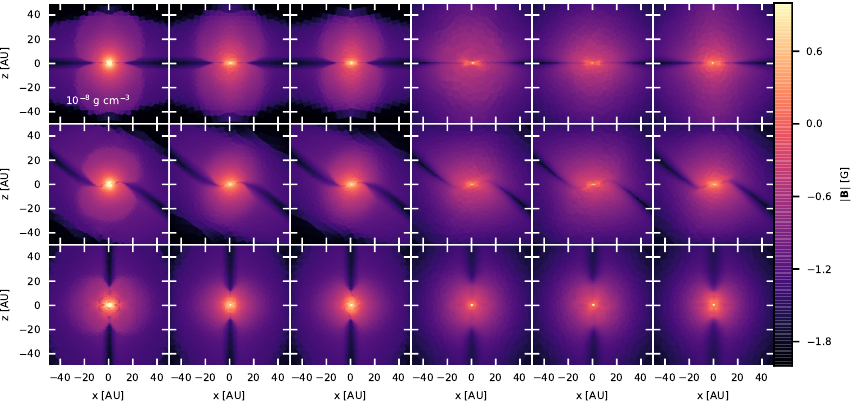}
    \caption{Slices of the absolute value of the magnetic field in the $xz$-plane at the time the simulation reaches $\rho_{\rm max} = 10^{-10}, \, 10^{-8} \, {\rm g} \,{\rm cm}^{-3}$ (top, bottom) for the first time. The magnetic field strength shows a much stronger gradient in the ideal MHD and $\zeta_i = 10^{-16}$ s$^{-1}$ cases, while it is more smooth if ambipolar diffusion is strong (17mrn models). The increase in central magnetic field strength at high temperatures (bottom plot) is seen in the latter.}
    \label{fig:babs_test}
\end{figure*}

\begin{figure*}
    \centering
    \includegraphics[width=1\linewidth]{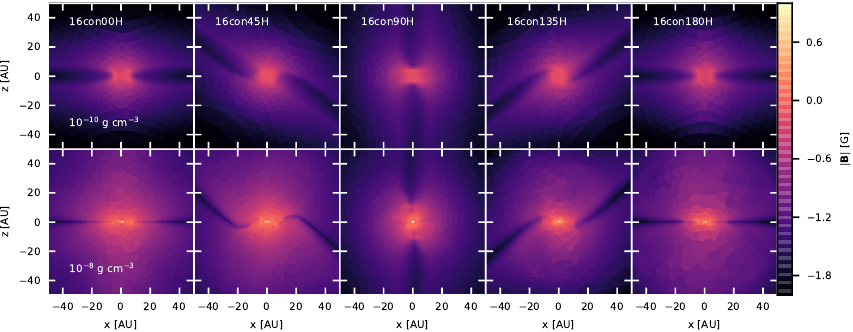}
    \caption{Vertical slices of the magnetic field strength for the Hall-dominated runs at $\rho_{\rm max} = 10^{-10}, \, 10^{-8} \, {\rm g} \,{\rm cm}^{-3}$ (top, bottom). We can see that the warping of the core/disk is connected to the warping of the magnetic field.}
    \label{fig:babs_hall}
\end{figure*}

\begin{figure*}
    \centering
    \includegraphics[width=0.26\linewidth]{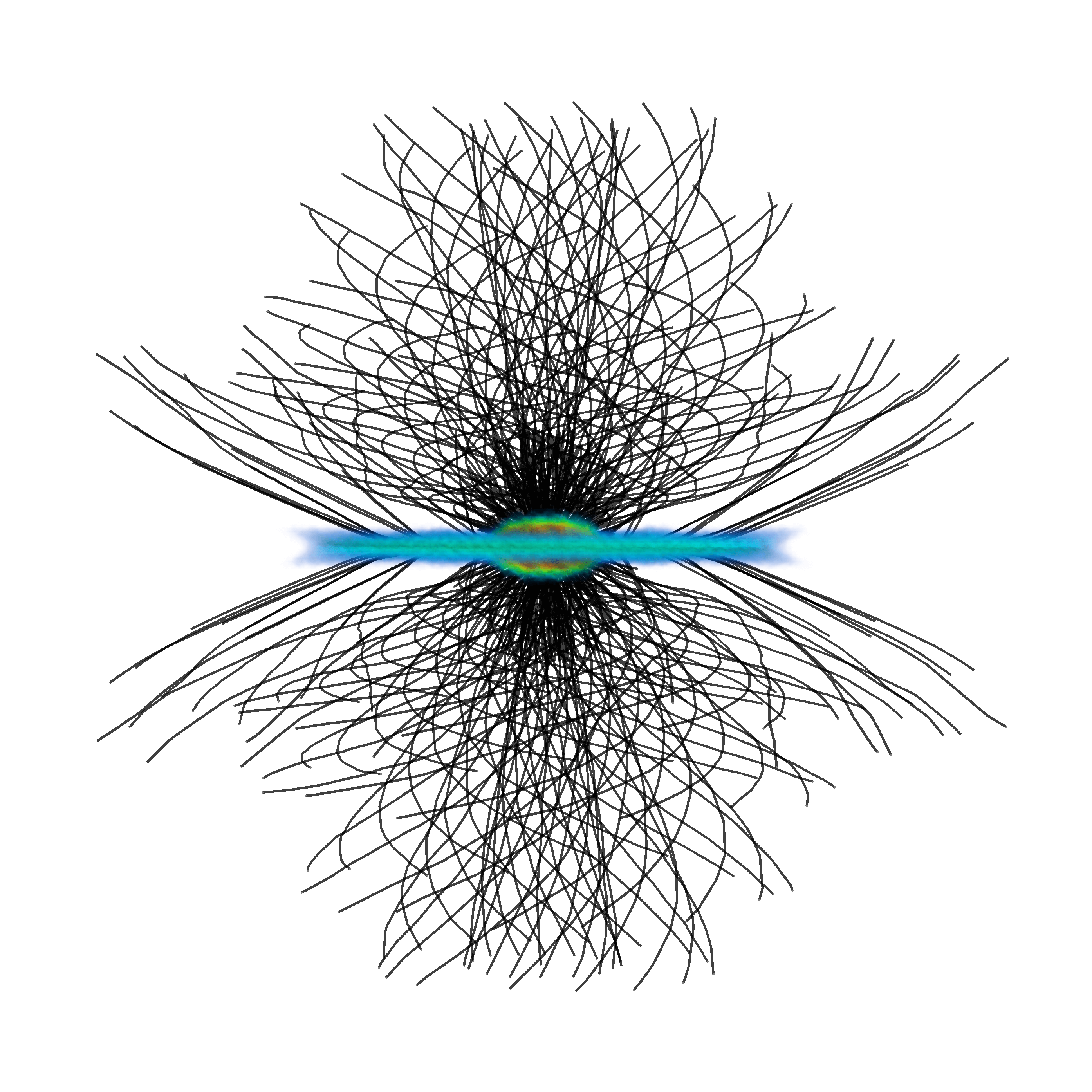}
    \includegraphics[width=0.26\linewidth]{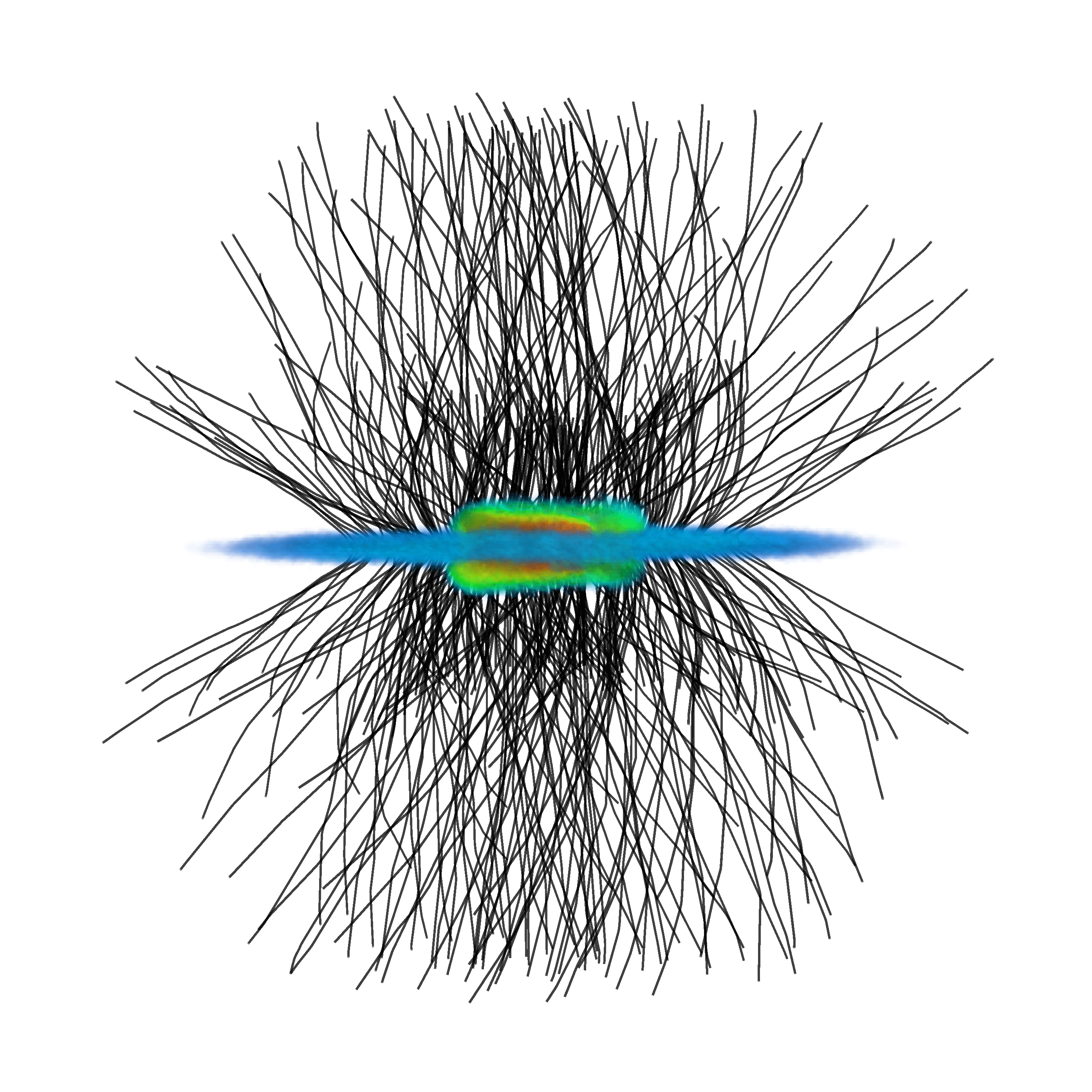}
    \includegraphics[width=0.26\linewidth]{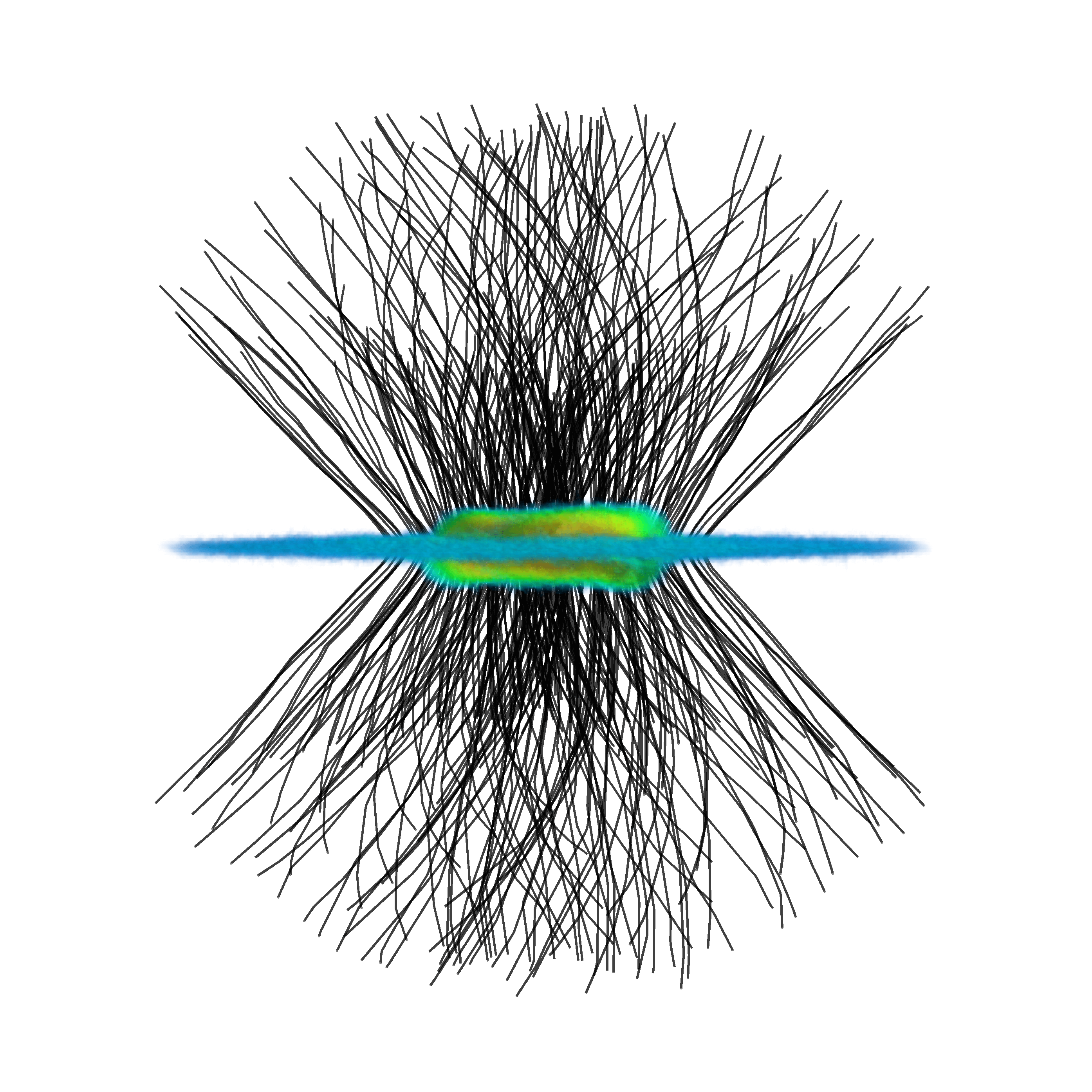}
    \vspace{1cm}
    \includegraphics[width=0.26\linewidth]{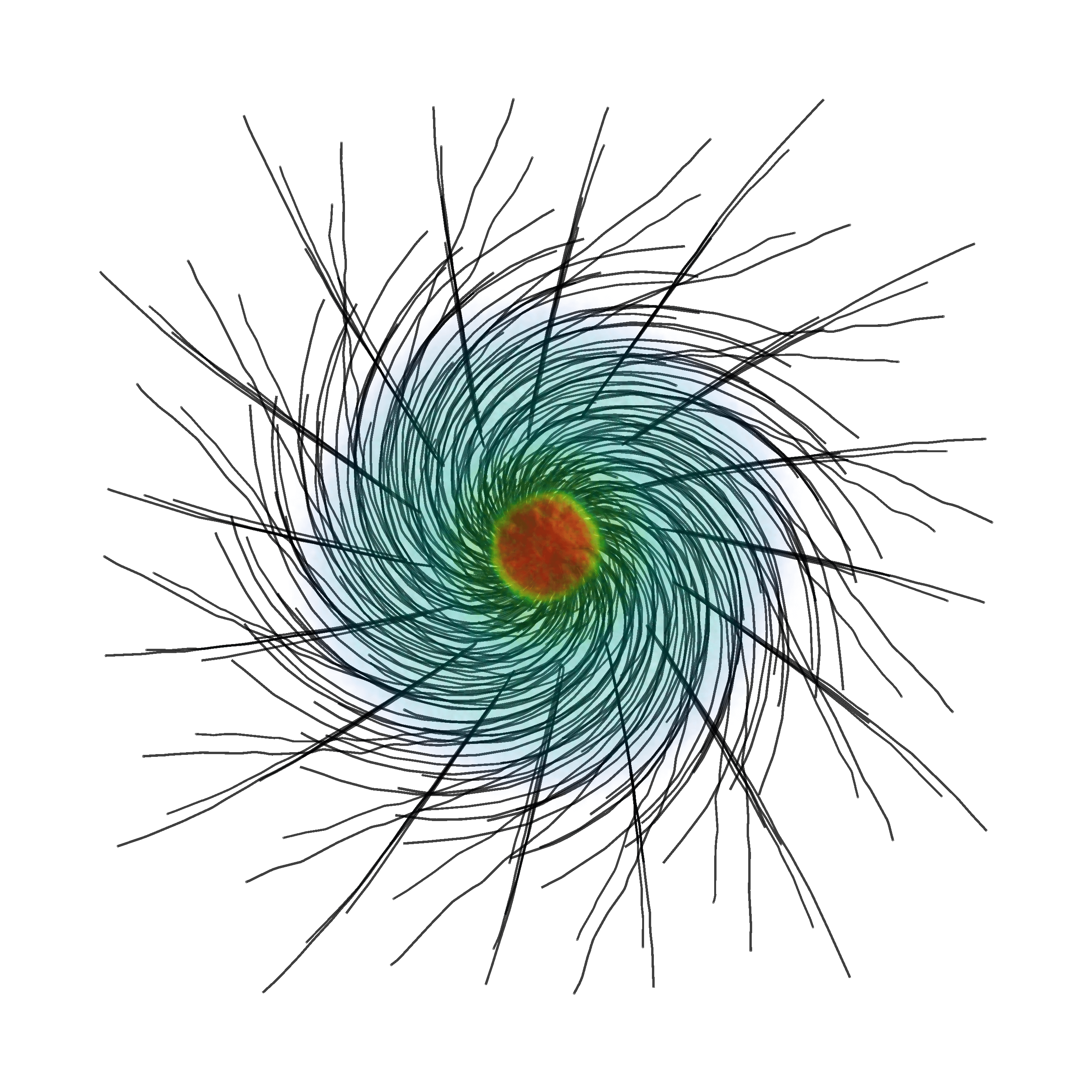}
    \includegraphics[width=0.26\linewidth]{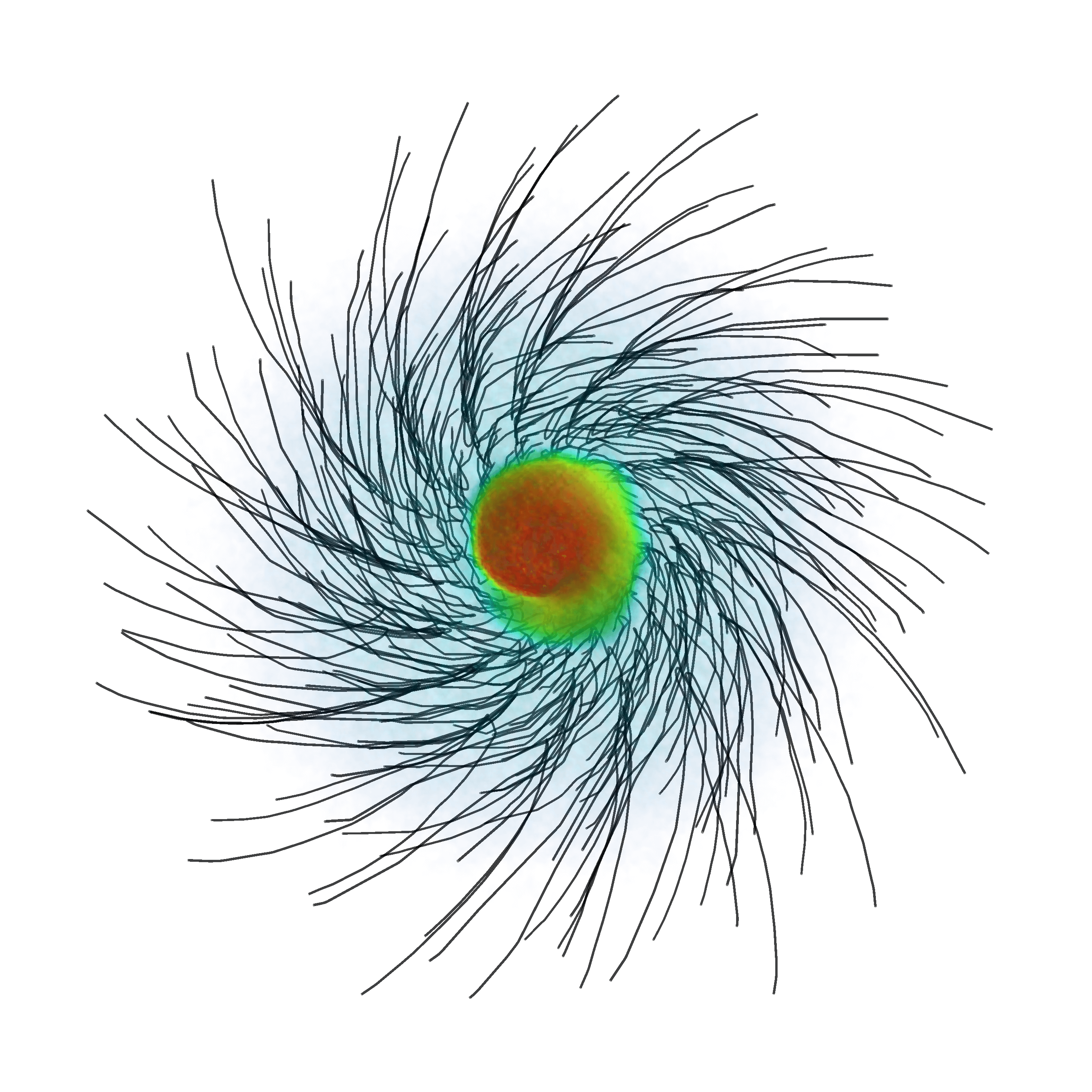}
    \includegraphics[width=0.26\linewidth]{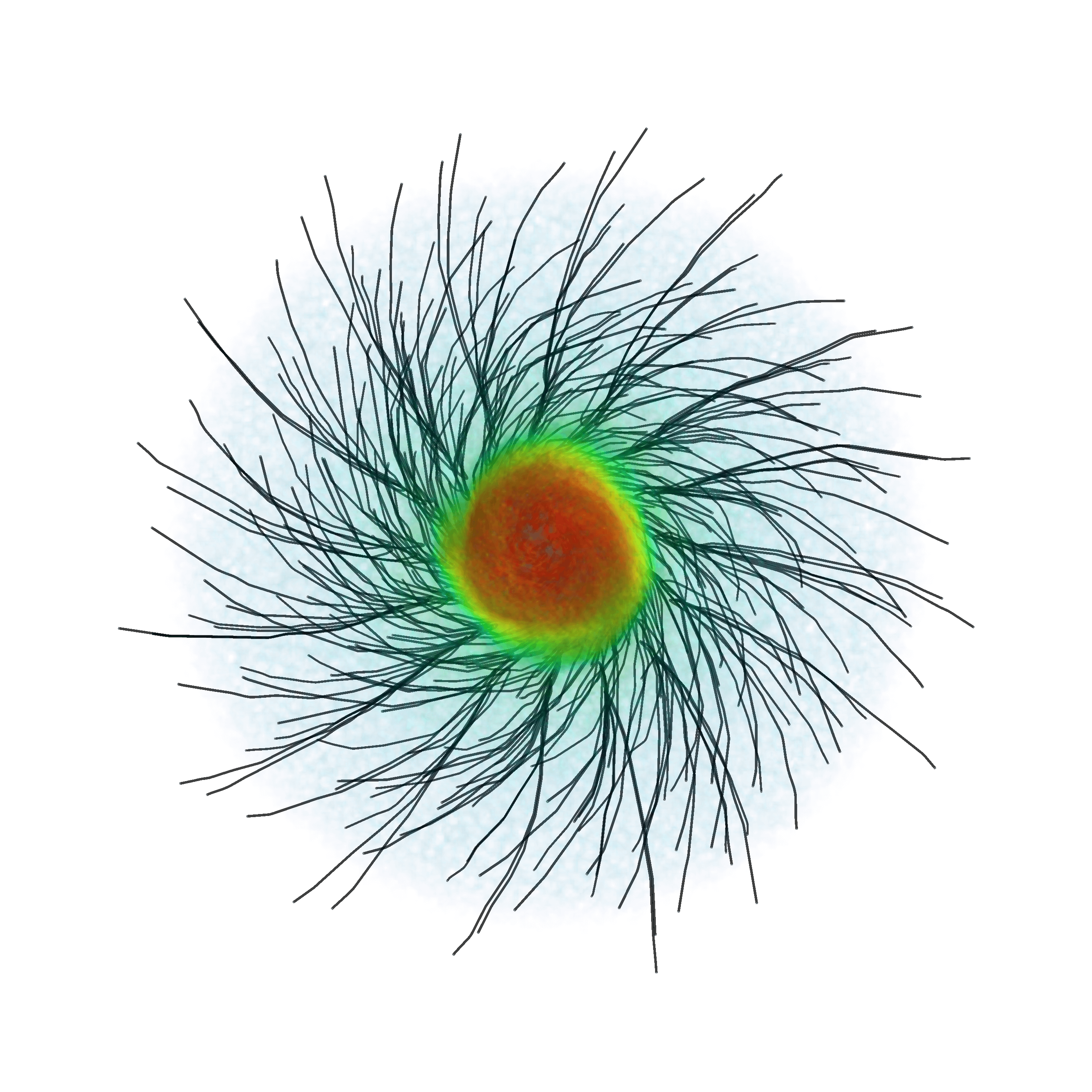}
    \includegraphics[width=0.26\linewidth]{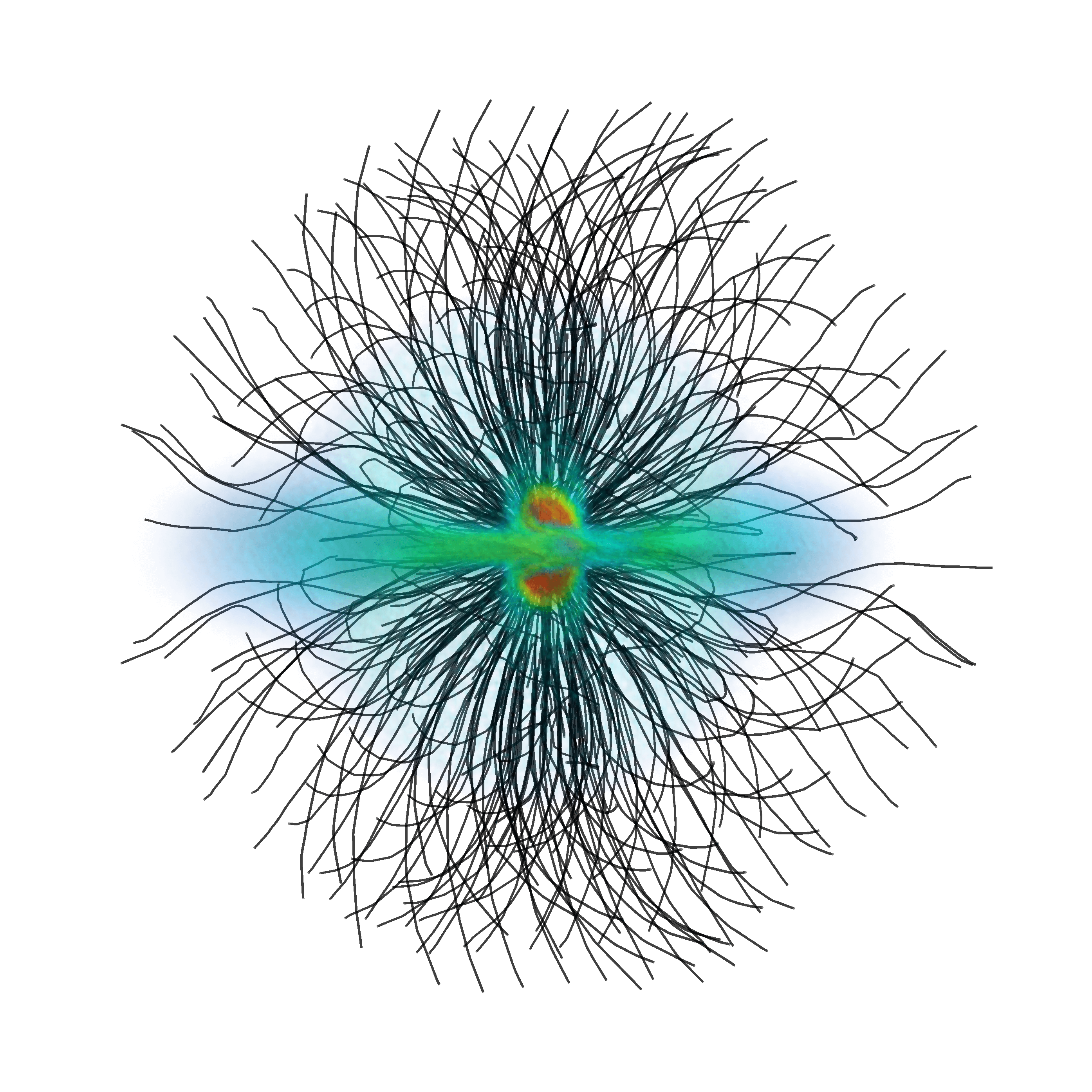}
    \includegraphics[width=0.26\linewidth]{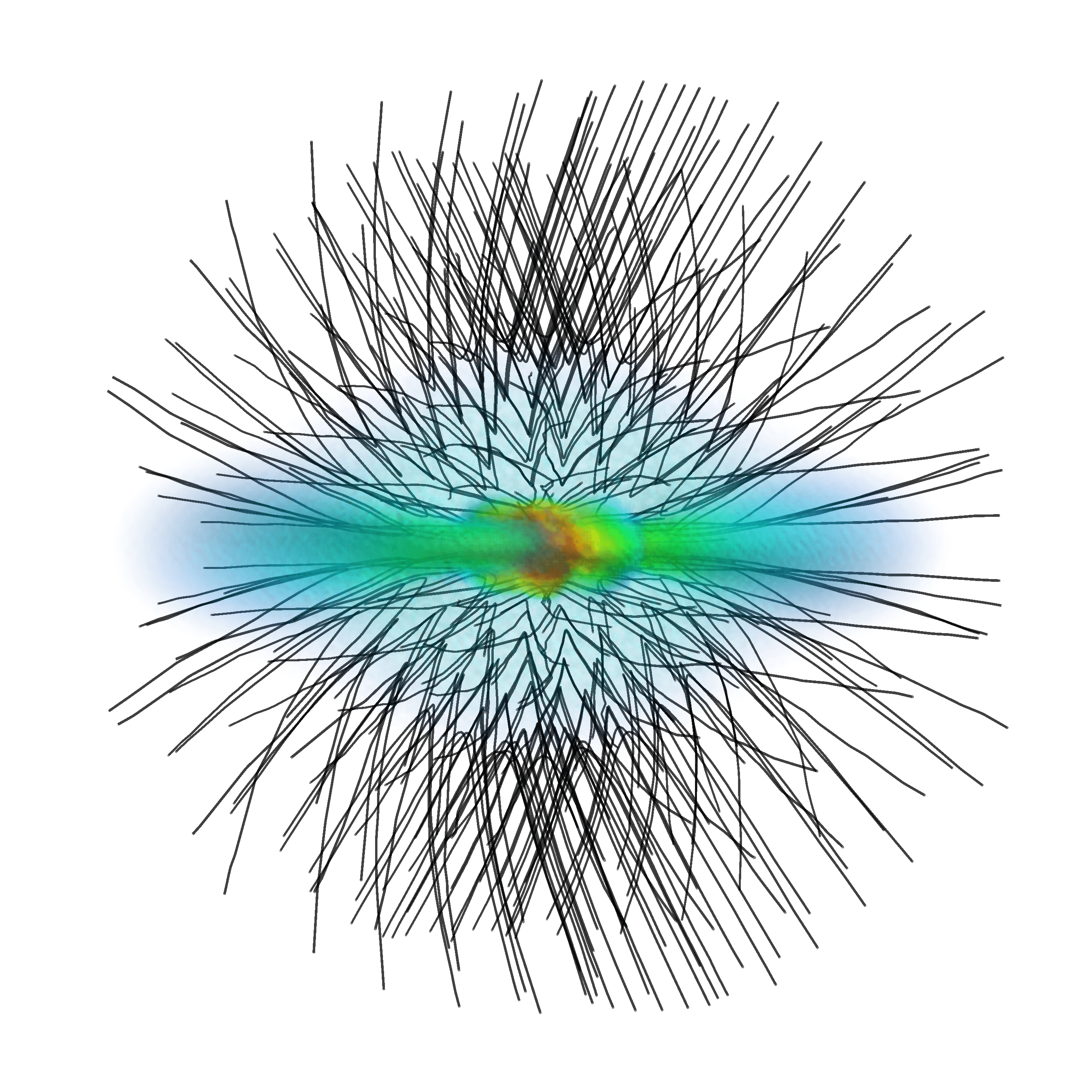}
    \includegraphics[width=0.26\linewidth]{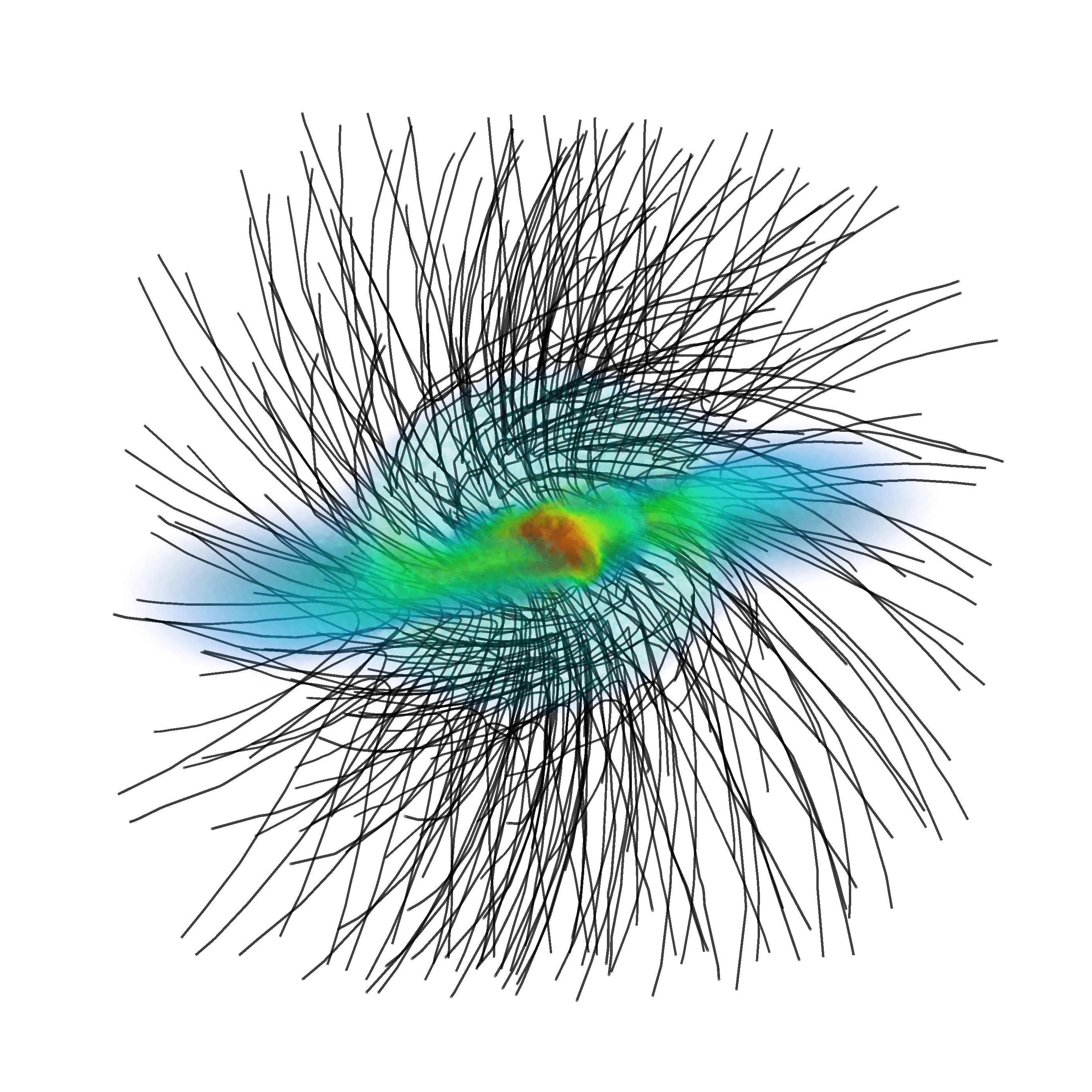}
    \includegraphics[width=0.26\linewidth]{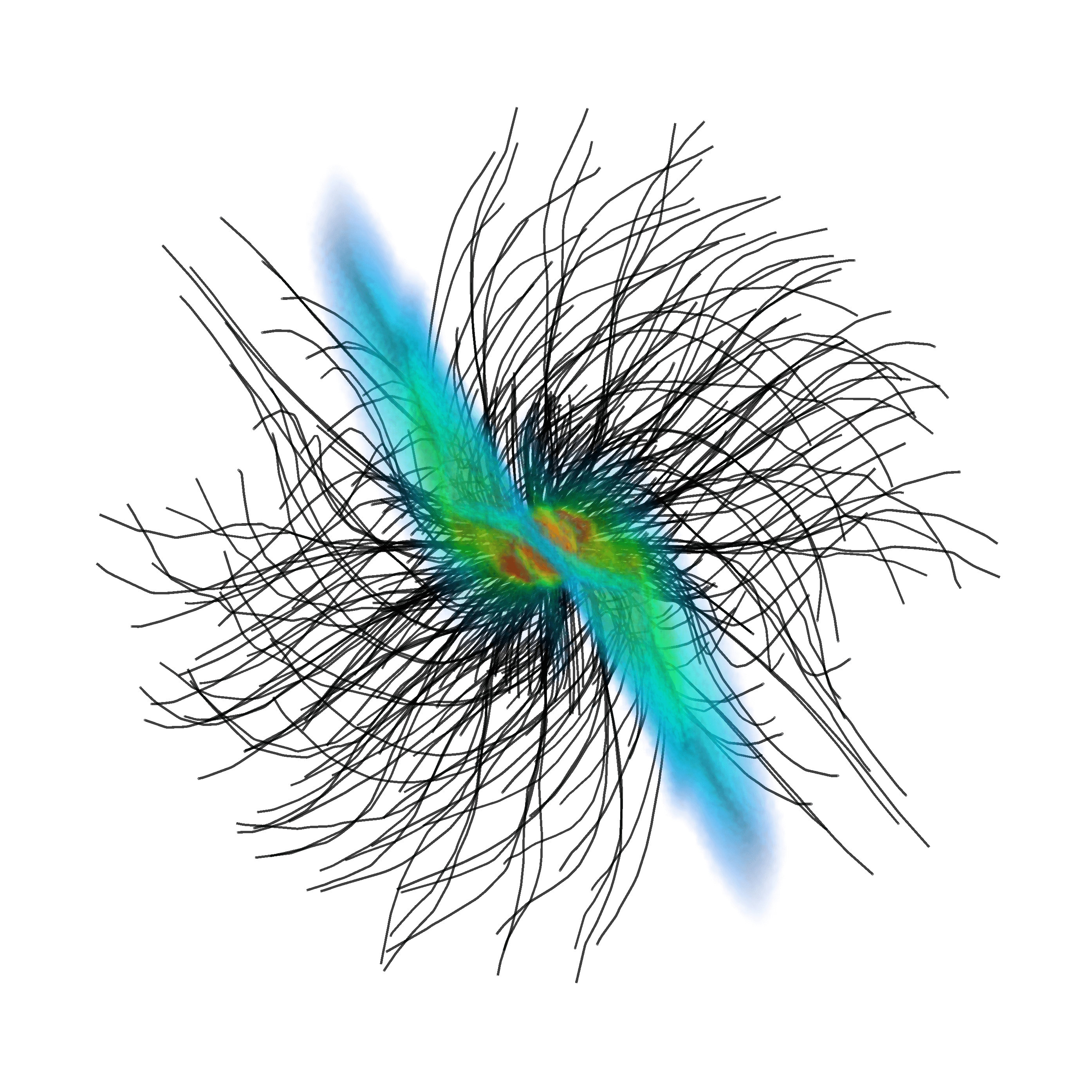}
    \includegraphics[width=0.26\linewidth]{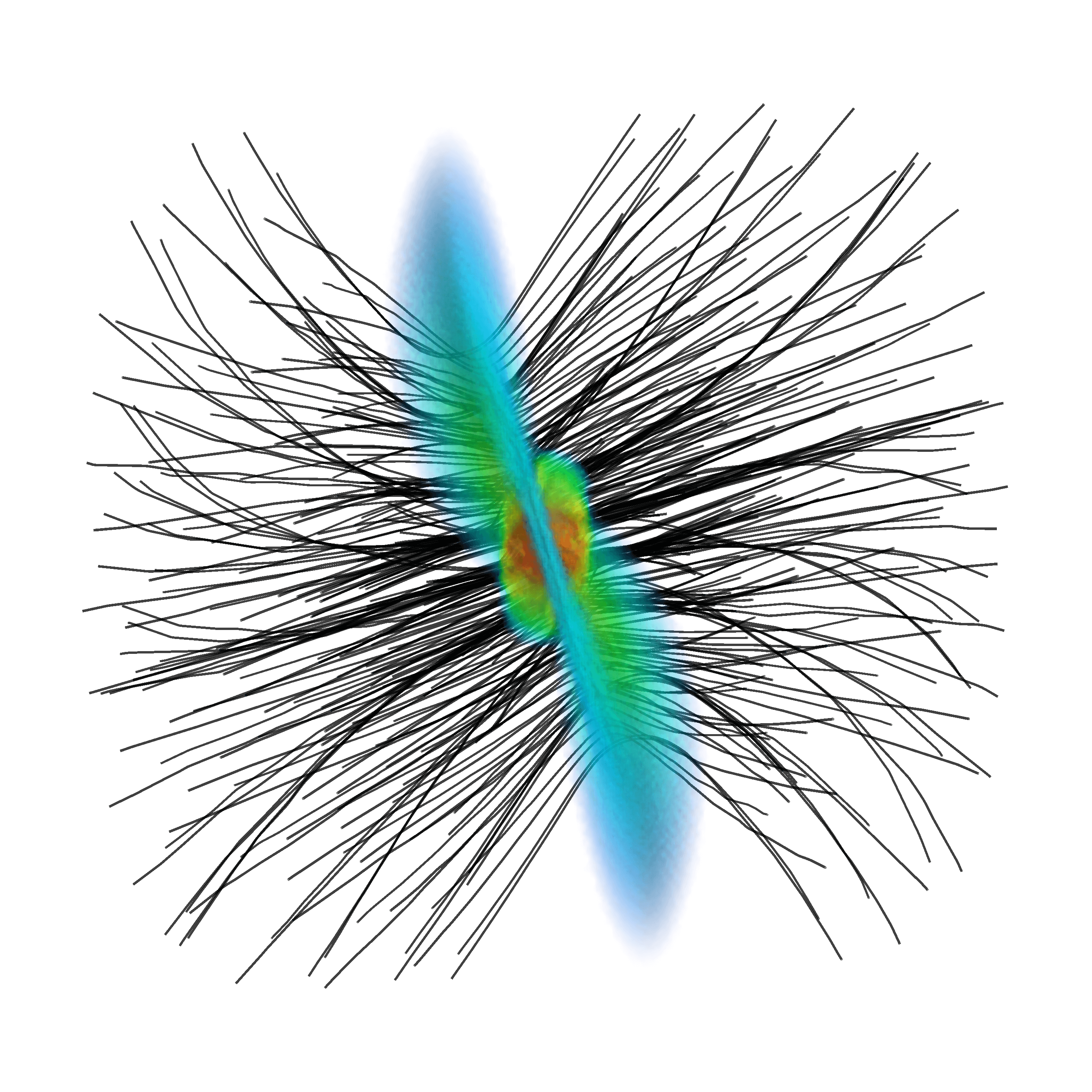}
    \includegraphics[width=0.26\linewidth]{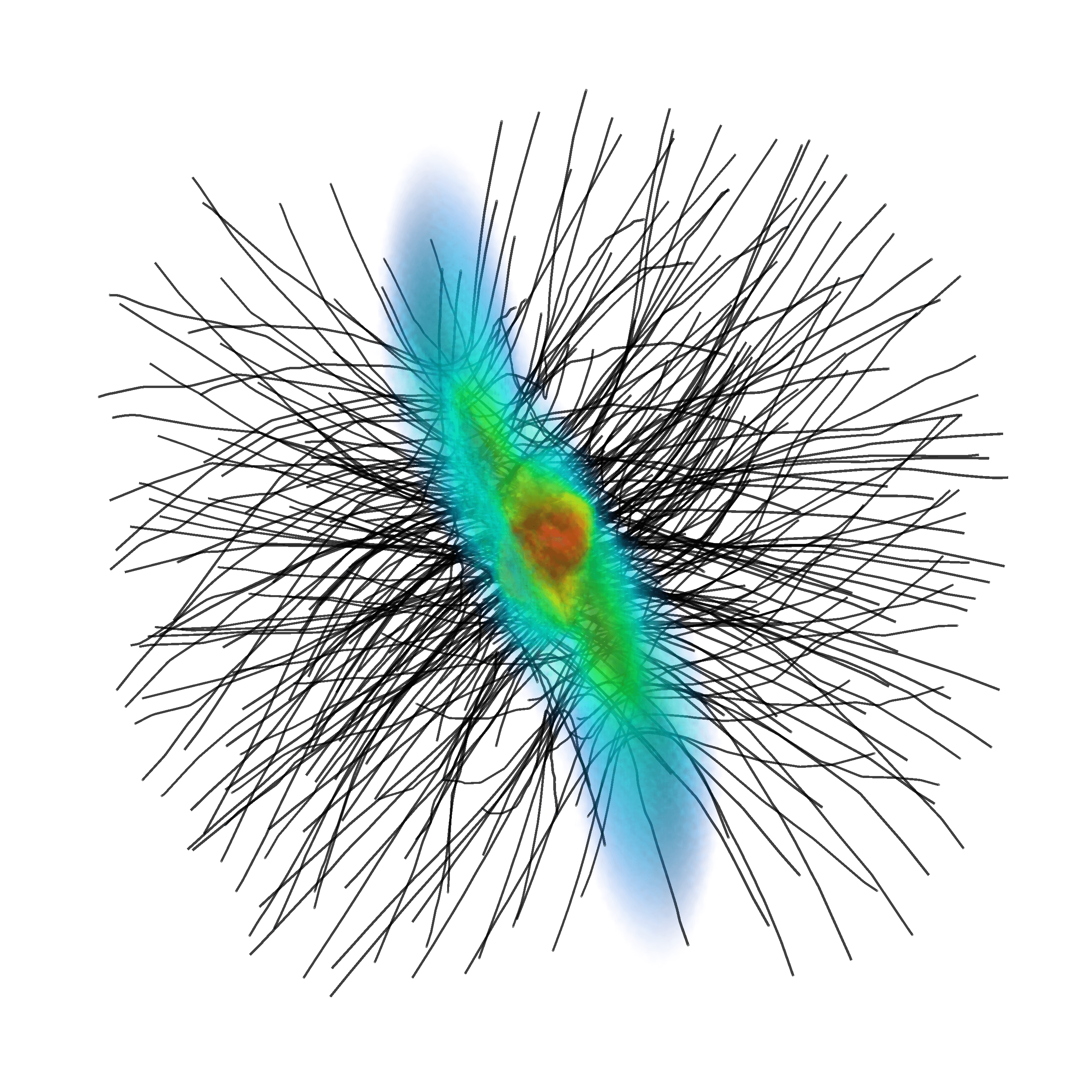}
    \caption{Visualization of the three dimensional structure of the central dense core/disk and the surrounding magnetic field $\rho_{\rm max} = 10^{-8}$ g cm$^{-3}$ for selected models: Higher densities in red, lower densities in blue; magnetic field lines as black tubes. The top 6 panels show the case of $00^\circ$ inclination, while on the bottom the more complex case of $90^\circ$ is displayed. In both cases, we have from left to right: Ideal MHD, 17mrnN, 16conH; the upper panels of each group show an edge-on, the lower panels a face-on view.}
    \label{fig:streamplots}
\end{figure*}

Figures~\ref{fig:babs_test} and \ref{fig:babs_hall} depict the absolute value of the magnetic field at different maximum densities in the simulation. At $\rho_{\rm max} = 10^{-10} \,{\rm g} \,{\rm cm}^{-3}$, the direction of the initial field has a clear impact on the magnetic field structure, and the morphology corresponds to those seen in a typical hourglass-shape of magnetic field lines (cf.~\ref{fig:streamplots} for a 3D view). This is consistent with our earlier observation that the orientation of the pseudo-disk, mainly supported by magnetic tension, is mostly determined by the large-scale magnetic field. There is already a clear reduction in the central (corresponding also to the maximal) magnetic field strength if non-ideal MHD is included (the left column shows ideal MHD). At this time, this is mostly due to ambipolar diffusion, as Ohmic resistivity is still very weak at these densities and the differences between simulations with and without the Hall effect are negligible compared to those between chemical models overall. However, the influence of the Hall effect is already barely visible as a twist in the magnetic field structure in all but the (anti-)aligned models. As the evolution progresses, differences between magnetic field models become even more pronounced, although a common feature is the enhancement in the central magnetic field caused by the recoupling of the magnetic field due to the rising temperature in the densest region. Overall, the magnetic field is substantially reduced in the 16con and 17mrn models with respect to ideal MHD and 16mrn. The diffusion works to smooth gradients between the first core and disk plane, and their surroundings, as the rise in field strength towards the center is much less pronounced. Outflows as seen in Figure~\ref{fig:rad_test} can be matched to regions of enhanced magnetic field, which matches the observation that the outflows are magnetically dominated (cf.~Figure \ref{fig:poynting_all} in Appendix~\ref{app:poyntingVsKineticFlux}). 

In Figure~\ref{fig:streamplots}, we aim to give an idea of the three-dimensional morphology of the magnetic field surrounding the disk by displaying the hydrostatic core and the streamlines of the magnetic field surrounding it. These plots are again turned as described in section~\ref{subsection:(pseudo-)disk} for the projections. In the 6 upper panels, displaying the $00^\circ$ case, one can get a visual impression for how the non-ideal effects lead to reduced azimuthal twisting of the magnetic field lines -- despite the fact that these disks rotate faster than in the ideal MHD case on the left. The strong Hall effect in 16con00H has been particularly effective in this regard. In the lower panels, showing simulations with $90^\circ$ initial misalignment, the magnetic field lines are also focused into the central core more strongly in ideal MHD than in the other two simulations. One can see here how the twisting due to the Hall effect works in terms of the magnetic field lines: While the 17mrn90N simulation still resembles an hourglass (although flipped onto its side), the structure is much more complex in 16con90H.

\section{Discussion and comparison to the literature}
\label{sec:discussion}
Due to our lack of radiative transfer and the resulting limitations of the accuracy of the temperature in the simulations, the diffusivity coefficients are potentially locally different from simulations including these effects. In particular, the sign of the Hall effect might differ, and its impact locally might thus be reversed from that in other calculations. Concerning precisely the Hall effect, however, the impact of using a barotropic equation of state has not been studied. A number of works have treated second core formation with non-ideal MHD and radiative transfer in the form of flux-limited diffusion \citep{wurster2021nonidealimpactsingle,vaytet2018protostellar,tsukamoto2017hall}, as well as additionally including moment-based methods \citep{mignon-rise2020hydrid}. Some use an approximate treatment \citep{xu2021formationII} or cooling along with a variable equation of state \citep{sadanari2023firststars}. Previous work has also established differences for a more complete temperature treatment, both concerning individual objects \citep[e.g.][]{tomida2013coreformation,prole2024popIII} as well as at a statistical level \citep{lebreuilly2024clumps}.

The number of studies including the Hall effect is in general still much smaller than those with just diffusion effects. To our knowledge this work presents the first 3D global simulations of stellar core formation with a non-SPH code. This is likely a result of the aforementioned numerical difficulties arising from the modelling of the dispersive whistler waves, which cause instabilities if not counteracted by sufficient diffusion. In SPH, this is usually sufficiently provided by the artificial resistivity that is already a requirement for ideal MHD, although this is likely dependent on the specific situation and implementation. Unfortunately, we were not able to run stable simulations that only include the Hall effect without any diffusive effects, such as those performed by \cite{wurster2021nonidealimpactsingle}, who actually conclude that the Hall effect is in their model (cosmic-ray ionization rate of $10^{-17}\, {\rm s}^{-1}$, constant grain size) the dominant non-ideal MHD process. Even though we have placed limits on the Hall coefficients, it still plays a significant role in the evolution of our simulated disks. However, it has recently been pointed out by \cite{hopkins2024hall} that in some situations the strength of the Hall effect in particular is likely overestimated in standard formulations of non-ideal MHD as relevant microphysics are neglected.

A substantial limitation in many studies with non-ideal MHD is the lack of evolution in the grain size distribution. As has been shown by many recent simulations and analytical considerations \citep[e.g.][]{tsukamoto2023grainsI,tsukamoto2024grainsII}, dust grain growth is significant on timescales of the evolution of protostellar disks, and has a large impact on the strength of non-ideal MHD effects and thereby the evolution of the magnetic field. 

Simulations of isolated molecular cloud cores suffer by construction from a lack of realistic turbulence. Even if \citep[e.g. in][]{wurster2020turbI, tu2024drod} a turbulent velocity spectrum is present in the initial conditions, this does not capture the full co-evolution of velocity, density and magnetic field, which would have preceded the formation of the cloud core. In models including accretion from turbulent larger scales \citep[e.g.][]{kuffmeier2017zoomin,heigl2024disk}, material that accretes onto the disk during its later evolution will generally not have the same axis of angular momentum and can change the axis of rotation substantially. 

In general, our model cannot give insights into the evolution of disk and outflow after second core formation, and our conclusions are therefore limited to very young stellar objects. Studies that instead focus on the long-term evolution of the disk by employing sink particles \citep[e.g.][]{zhao2020hall,zhao2021interplay,tu2024drod,lebreuilly2024clumps} or limiting the resolution in the core \citep{machida_basu2019twothousand} can provide insights into the long-term behaviour of the disk.

Lastly, numerical resolution has been shown to affect the evolution and morphology of protostellar disks even in the presence of physical diffusion in the form of non-ideal MHD \citep{wurster2022resolution}. We have performed a comparison simulation to 16con00H which is identical to the on presented in the body of this work except for its reduced target mass resolution of $10^{-6}$ M$_\odot$. We find that the difference in time to collapse to the second core is <0.02 kyr (substantially less than the variation between most of the chemical models), with an earlier collapse at low-resolution, matching the findings of \cite{wurster2022resolution}. The disk also develops gravitational instability and the overall morphology is very similar.

\section{Summary and conclusion}
\label{sec:summary}
We have performed a large suite of high-resolution ($3.33 \times 10^{-7}\, {\rm M}_\odot$, medium cell mass) simulations representing the idealized collapse of molecular cloud cores (mass of $M =1\, {\rm M}_\odot$, magnetization corresponding to $\mu = 5.0$ and rotational energy of $\beta_{\rm r} = 0.005$) to stellar cores with a range of different models for the evolution of the magnetic field (considering both ideal and non-ideal MHD) and initial relative orientations of rotation axis and magnetic field. We find that both of these model parameters have a substantial effect on the formation of the first core and the subsequent (non-)emergence of a rotationally supported disk. Our main conclusions are as follows: 

\begin{enumerate}
    
\item We confirm earlier results that at this level of initial magnetization, rotationally supported disks do not form in the ideal MHD case, irrespective of initial geometry. This is due to the efficient removal of angular momentum through magnetic braking. In all models, the pseudo-disk, which forms early in the evolution, is perpendicular to the initial field orientation.

\item Independent of field orientation, the inclusion of non-ideal MHD increases the specific angular momentum of the first core through weakening the magnetic braking, as the magnetic field strength is reduced by orders of magnitude. This is generally correlated to a longer time spent in the first core phase, but the $90^\circ$ case serves as an exception, likely because the larger magnetic support of ideal MHD dominates the low rotational support of all simulations with this orientation. The simulations with this initial condition are distinct in general, as they all show early collapse and a general morphology that is quite different from the $45^\circ$ case, which is still largely similar to the `basic' setup of $00^\circ$.

\item Ambipolar diffusion is the dominant effect in the early collapse. The importance of the Hall effect is significantly greater for a constant grain size than for an MRN size distribution. In the former case, the Hall effect can be strong enough that models with larger diffusion end up with less specific angular momentum if the initial magnetic field and rotation axis are (close to) aligned, as 16con00H shows a higher value than 17mrn00H (and 16con45H than 17mrn45H), while 16con00N (16con45N) does not. On the other hand, 16con180H and 16con135H have reduced rotational speeds compared to 16con00H and 16con45H, respectively. 

\item Outflows are produced in all ideal MHD simulations, and the mass of outflowing material is generally reduced by the presence of diffusion, which can is some cases even lead to the absence of first-core outflows. Outflows have broader opening angles and are larger in non-ideal MHD, but speeds are lower. In almost all cases, the strength of outflows decreases as the initial offset angle is increased from (anti-)aligned to $90^\circ$. The Hall effect can lead to both more and less prominent outflows, depending on the geometry, and even provide new outflow channels. 

\end{enumerate}

While we have in this work focused on idealized setups, providing a good laboratory to isolate the influence of individual physical effects (grain size distribution, ionization rate, the Hall effect, initial geometry), we aim to use the knowledge gained from the experiments to further  our understanding of more complex situations. {\small AREPO} is a very flexible code that, as we have shown, can handle the large range of spatial and temporal scales present in the modelling of star formation without issue. 

Our next steps are to work on two of the limitations of our models presented here: The lack of turbulence from the larger scale and the simplified temperature treatment. We are therefore preparing to run larger scale simulations that start from ISM boxes with turbulence driven by supernova explosions in which we resolve the full formation process of protostars, including the formation of the second core. Similarly, the plan is to develop a version of flux-limited diffusion on an unstructured mesh and include it in future simulations of cloud core collapse with {\small AREPO}.

\section*{Acknowledgements}
The authors acknowledge helpful discussions with Michael K\"uffmeier and Tommaso Grassi. OZ acknowledges support from the ITC Postdoctoral Fellowship. SW acknowledges the Deutsche Forschungsgemeinschaft (DFG) for funding through the SFB 1601 ``Habitats of massive stars across cosmic time'' (sub-project A5).

\section*{Data Availability}
The data underlying this paper will be shared upon reasonable request to the corresponding author.

\bibliographystyle{mnras}
\bibliography{bibliography.bib}

\begin{thebibliography}{}
\makeatletter
\relax
\def\mn@urlcharsother{\let\do\@makeother \do\$\do\&\do\#\do\^\do\_\do\%\do\~}
\def\mn@doi{\begingroup\mn@urlcharsother \@ifnextchar [ {\mn@doi@}
  {\mn@doi@[]}}
\def\mn@doi@[#1]#2{\def\@tempa{#1}\ifx\@tempa\@empty \href
  {http://dx.doi.org/#2} {doi:#2}\else \href {http://dx.doi.org/#2} {#1}\fi
  \endgroup}
\def\mn@eprint#1#2{\mn@eprint@#1:#2::\@nil}
\def\mn@eprint@arXiv#1{\href {http://arxiv.org/abs/#1} {{\tt arXiv:#1}}}
\def\mn@eprint@dblp#1{\href {http://dblp.uni-trier.de/rec/bibtex/#1.xml}
  {dblp:#1}}
\def\mn@eprint@#1:#2:#3:#4\@nil{\def\@tempa {#1}\def\@tempb {#2}\def\@tempc
  {#3}\ifx \@tempc \@empty \let \@tempc \@tempb \let \@tempb \@tempa \fi \ifx
  \@tempb \@empty \def\@tempb {arXiv}\fi \@ifundefined
  {mn@eprint@\@tempb}{\@tempb:\@tempc}{\expandafter \expandafter \csname
  mn@eprint@\@tempb\endcsname \expandafter{\@tempc}}}

\bibitem[\protect\citeauthoryear{{Ahmad}, {Gonz{\'a}lez}, {Hennebelle}  \&
  {Commer{\c{c}}on}}{{Ahmad} et~al.}{2023}]{ahmad2023protostar}
{Ahmad} A.,  {Gonz{\'a}lez} M.,  {Hennebelle} P.,   {Commer{\c{c}}on} B.,
  2023, \mn@doi [\aap] {10.1051/0004-6361/202346711}, \href
  {https://ui.adsabs.harvard.edu/abs/2023A&A...680A..23A} {680, A23}

\bibitem[\protect\citeauthoryear{{Allen}, {Li}  \& {Shu}}{{Allen}
  et~al.}{2003}]{allen2003braking}
{Allen} A.,  {Li} Z.-Y.,   {Shu} F.~H.,  2003, \mn@doi [\apj] {10.1086/379243},
  \href {https://ui.adsabs.harvard.edu/abs/2003ApJ...599..363A} {599, 363}

\bibitem[\protect\citeauthoryear{{Alves}, {Caselli}, {Girart}, {Segura-Cox},
  {Franco}, {Schmiedeke}  \& {Zhao}}{{Alves} et~al.}{2019}]{alves2019ybp}
{Alves} F.~O.,  {Caselli} P.,  {Girart} J.~M.,  {Segura-Cox} D.,  {Franco}
  G.~A.~P.,  {Schmiedeke} A.,   {Zhao} B.,  2019, \mn@doi [Science]
  {10.1126/science.aaw3491}, \href
  {https://ui.adsabs.harvard.edu/abs/2019Sci...366...90A} {366, 90}

\bibitem[\protect\citeauthoryear{{Bai}}{{Bai}}{2014}]{bai2014hall}
{Bai} X.-N.,  2014, \mn@doi [\apj] {10.1088/0004-637X/791/2/137}, \href
  {https://ui.adsabs.harvard.edu/abs/2014ApJ...791..137B} {791, 137}

\bibitem[\protect\citeauthoryear{{Bai} \& {Stone}}{{Bai} \&
  {Stone}}{2011}]{bai2011ambipolar}
{Bai} X.-N.,  {Stone} J.~M.,  2011, \mn@doi [\apj]
  {10.1088/0004-637X/736/2/144}, \href
  {https://ui.adsabs.harvard.edu/abs/2011ApJ...736..144B} {736, 144}

\bibitem[\protect\citeauthoryear{{Bate}, {Tricco}  \& {Price}}{{Bate}
  et~al.}{2014}]{bate2014collapse}
{Bate} M.~R.,  {Tricco} T.~S.,   {Price} D.~J.,  2014, \mn@doi [\mnras]
  {10.1093/mnras/stt1865}, \href
  {https://ui.adsabs.harvard.edu/abs/2014MNRAS.437...77B} {437, 77}

\bibitem[\protect\citeauthoryear{{Favre} et~al.,}{{Favre}
  et~al.}{2018}]{Favre2018}
{Favre} C.,  et~al., 2018, \mn@doi [\apj] {10.3847/1538-4357/aabfd4}, \href
  {https://ui.adsabs.harvard.edu/abs/2018ApJ...859..136F} {859, 136}

\bibitem[\protect\citeauthoryear{{Fitz Axen}, {Offner}, {Hopkins}, {Krumholz}
  \& {Grudic}}{{Fitz Axen} et~al.}{2024}]{Axen2024}
{Fitz Axen} M.,  {Offner} S.,  {Hopkins} P.~F.,  {Krumholz} M.~R.,   {Grudic}
  M.~Y.,  2024, \mn@doi [arXiv e-prints] {10.48550/arXiv.2407.17597}, \href
  {https://ui.adsabs.harvard.edu/abs/2024arXiv240717597F} {p. arXiv:2407.17597}

\bibitem[\protect\citeauthoryear{{Flores} et~al.,}{{Flores}
  et~al.}{2023}]{flores2023streamer}
{Flores} C.,  et~al., 2023, \mn@doi [\apj] {10.3847/1538-4357/acf7c1}, \href
  {https://ui.adsabs.harvard.edu/abs/2023ApJ...958...98F} {958, 98}

\bibitem[\protect\citeauthoryear{{Galli} \& {Shu}}{{Galli} \&
  {Shu}}{1993a}]{Galli&Shu1993I}
{Galli} D.,  {Shu} F.~H.,  1993a, \mn@doi [\apj] {10.1086/173305}, \href
  {https://ui.adsabs.harvard.edu/abs/1993ApJ...417..220G} {417, 220}

\bibitem[\protect\citeauthoryear{{Galli} \& {Shu}}{{Galli} \&
  {Shu}}{1993b}]{Galli&Shu1993II}
{Galli} D.,  {Shu} F.~H.,  1993b, \mn@doi [\apj] {10.1086/173306}, \href
  {https://ui.adsabs.harvard.edu/abs/1993ApJ...417..243G} {417, 243}

\bibitem[\protect\citeauthoryear{{Hanasz}, {Strong}  \& {Girichidis}}{{Hanasz}
  et~al.}{2021}]{hanasz2021cosmicrays}
{Hanasz} M.,  {Strong} A.~W.,   {Girichidis} P.,  2021, \mn@doi [Living Reviews
  in Computational Astrophysics] {10.1007/s41115-021-00011-1}, \href
  {https://ui.adsabs.harvard.edu/abs/2021LRCA....7....2H} {7, 2}

\bibitem[\protect\citeauthoryear{{Heigl}, {Hoemann}  \& {Burkert}}{{Heigl}
  et~al.}{2024}]{heigl2024disk}
{Heigl} S.,  {Hoemann} E.,   {Burkert} A.,  2024, \mn@doi [arXiv e-prints]
  {10.48550/arXiv.2401.03779}, \href
  {https://ui.adsabs.harvard.edu/abs/2024arXiv240103779H} {p. arXiv:2401.03779}

\bibitem[\protect\citeauthoryear{{Hennebelle} \& {Ciardi}}{{Hennebelle} \&
  {Ciardi}}{2009}]{hennebelle2009misalignment}
{Hennebelle} P.,  {Ciardi} A.,  2009, \mn@doi [\aap]
  {10.1051/0004-6361/200913008}, \href
  {https://ui.adsabs.harvard.edu/abs/2009A&A...506L..29H} {506, L29}

\bibitem[\protect\citeauthoryear{{Hennebelle} \& {Fromang}}{{Hennebelle} \&
  {Fromang}}{2008}]{hennebelle2008magnetic}
{Hennebelle} P.,  {Fromang} S.,  2008, \mn@doi [\aap]
  {10.1051/0004-6361:20078309}, \href
  {https://ui.adsabs.harvard.edu/abs/2008A&A...477....9H} {477, 9}

\bibitem[\protect\citeauthoryear{{Hirano}, {Tsukamoto}, {Basu}  \&
  {Machida}}{{Hirano} et~al.}{2020}]{Hirano2020misalignment}
{Hirano} S.,  {Tsukamoto} Y.,  {Basu} S.,   {Machida} M.~N.,  2020, \mn@doi
  [\apj] {10.3847/1538-4357/ab9f9d}, \href
  {https://ui.adsabs.harvard.edu/abs/2020ApJ...898..118H} {898, 118}

\bibitem[\protect\citeauthoryear{{Hopkins}}{{Hopkins}}{2017}]{hopkins2017diffusion}
{Hopkins} P.~F.,  2017, \mn@doi [\mnras] {10.1093/mnras/stw3306}, \href
  {https://ui.adsabs.harvard.edu/abs/2017MNRAS.466.3387H} {466, 3387}

\bibitem[\protect\citeauthoryear{{Hopkins}, {Squire}, {Skalidis}  \&
  {Soliman}}{{Hopkins} et~al.}{2024}]{hopkins2024hall}
{Hopkins} P.~F.,  {Squire} J.,  {Skalidis} R.,   {Soliman} N.~H.,  2024,
  \mn@doi [arXiv e-prints] {10.48550/arXiv.2405.06026}, \href
  {https://ui.adsabs.harvard.edu/abs/2024arXiv240506026H} {p. arXiv:2405.06026}

\bibitem[\protect\citeauthoryear{{Hsieh} et~al.,}{{Hsieh}
  et~al.}{2023}]{hsieh2023streamer}
{Hsieh} T.~H.,  et~al., 2023, \mn@doi [\aap] {10.1051/0004-6361/202244183},
  \href {https://ui.adsabs.harvard.edu/abs/2023A&A...669A.137H} {669, A137}

\bibitem[\protect\citeauthoryear{{Joos}, {Hennebelle}  \& {Ciardi}}{{Joos}
  et~al.}{2012}]{joos2012misalignment}
{Joos} M.,  {Hennebelle} P.,   {Ciardi} A.,  2012, \mn@doi [\aap]
  {10.1051/0004-6361/201118730}, \href
  {https://ui.adsabs.harvard.edu/abs/2012A&A...543A.128J} {543, A128}

\bibitem[\protect\citeauthoryear{{Kuffmeier}, {Haugb{\o}lle}  \&
  {Nordlund}}{{Kuffmeier} et~al.}{2017}]{kuffmeier2017zoomin}
{Kuffmeier} M.,  {Haugb{\o}lle} T.,   {Nordlund} {\r{A}}.,  2017, \mn@doi
  [\apj] {10.3847/1538-4357/aa7c64}, \href
  {https://ui.adsabs.harvard.edu/abs/2017ApJ...846....7K} {846, 7}

\bibitem[\protect\citeauthoryear{{Larson}}{{Larson}}{1969}]{larson1969}
{Larson} R.~B.,  1969, \mn@doi [\mnras] {10.1093/mnras/145.3.271}, \href
  {https://ui.adsabs.harvard.edu/abs/1969MNRAS.145..271L} {145, 271}

\bibitem[\protect\citeauthoryear{{Lebreuilly} et~al.,}{{Lebreuilly}
  et~al.}{2024}]{lebreuilly2024clumps}
{Lebreuilly} U.,  et~al., 2024, \mn@doi [\aap] {10.1051/0004-6361/202346558},
  \href {https://ui.adsabs.harvard.edu/abs/2024A&A...682A..30L} {682, A30}

\bibitem[\protect\citeauthoryear{{Lesur}, {Kunz}  \& {Fromang}}{{Lesur}
  et~al.}{2014}]{lesur2014thanatology}
{Lesur} G.,  {Kunz} M.~W.,   {Fromang} S.,  2014, \mn@doi [\aap]
  {10.1051/0004-6361/201423660}, \href
  {https://ui.adsabs.harvard.edu/abs/2014A&A...566A..56L} {566, A56}

\bibitem[\protect\citeauthoryear{{Li}, {Krasnopolsky}  \& {Shang}}{{Li}
  et~al.}{2011}]{li2011non}
{Li} Z.-Y.,  {Krasnopolsky} R.,   {Shang} H.,  2011, \mn@doi [\apj]
  {10.1088/0004-637X/738/2/180}, \href
  {https://ui.adsabs.harvard.edu/abs/2011ApJ...738..180L} {738, 180}

\bibitem[\protect\citeauthoryear{{Machida} \& {Basu}}{{Machida} \&
  {Basu}}{2019}]{machida_basu2019twothousand}
{Machida} M.~N.,  {Basu} S.,  2019, \mn@doi [\apj] {10.3847/1538-4357/ab18a7},
  \href {https://ui.adsabs.harvard.edu/abs/2019ApJ...876..149M} {876, 149}

\bibitem[\protect\citeauthoryear{{Machida} \& {Matsumoto}}{{Machida} \&
  {Matsumoto}}{2011}]{machida2011circumstellar}
{Machida} M.~N.,  {Matsumoto} T.,  2011, \mn@doi [\mnras]
  {10.1111/j.1365-2966.2011.18349.x}, \href
  {https://ui.adsabs.harvard.edu/abs/2011MNRAS.413.2767M} {413, 2767}

\bibitem[\protect\citeauthoryear{{Machida}, {Inutsuka}  \&
  {Matsumoto}}{{Machida} et~al.}{2014}]{machida2014sinks}
{Machida} M.~N.,  {Inutsuka} S.-i.,   {Matsumoto} T.,  2014, \mn@doi [\mnras]
  {10.1093/mnras/stt2343}, \href
  {https://ui.adsabs.harvard.edu/abs/2014MNRAS.438.2278M} {438, 2278}

\bibitem[\protect\citeauthoryear{{Machida}, {Hirano}  \& {Kitta}}{{Machida}
  et~al.}{2020}]{machida2020misalignment}
{Machida} M.~N.,  {Hirano} S.,   {Kitta} H.,  2020, \mn@doi [\mnras]
  {10.1093/mnras/stz3159}, \href
  {https://ui.adsabs.harvard.edu/abs/2020MNRAS.491.2180M} {491, 2180}

\bibitem[\protect\citeauthoryear{{Marchand}, {Commer{\c{c}}on}  \&
  {Chabrier}}{{Marchand} et~al.}{2018}]{marchand2018hall}
{Marchand} P.,  {Commer{\c{c}}on} B.,   {Chabrier} G.,  2018, \mn@doi [\aap]
  {10.1051/0004-6361/201832907}, \href
  {https://ui.adsabs.harvard.edu/abs/2018A&A...619A..37M} {619, A37}

\bibitem[\protect\citeauthoryear{{Masson}, {Teyssier}, {Mulet-Marquis},
  {Hennebelle}  \& {Chabrier}}{{Masson} et~al.}{2012}]{masson2012ambipolar}
{Masson} J.,  {Teyssier} R.,  {Mulet-Marquis} C.,  {Hennebelle} P.,
  {Chabrier} G.,  2012, \mn@doi [\apjs] {10.1088/0067-0049/201/2/24}, \href
  {https://ui.adsabs.harvard.edu/abs/2012ApJS..201...24M} {201, 24}

\bibitem[\protect\citeauthoryear{{Mathis}, {Rumpl}  \& {Nordsieck}}{{Mathis}
  et~al.}{1977}]{mathis1977mrn}
{Mathis} J.~S.,  {Rumpl} W.,   {Nordsieck} K.~H.,  1977, \mn@doi [\apj]
  {10.1086/155591}, \href
  {https://ui.adsabs.harvard.edu/abs/1977ApJ...217..425M} {217, 425}

\bibitem[\protect\citeauthoryear{{Matsumoto} \& {Tomisaka}}{{Matsumoto} \&
  {Tomisaka}}{2004}]{matsumoto2014direction}
{Matsumoto} T.,  {Tomisaka} K.,  2004, \mn@doi [\apj] {10.1086/424897}, \href
  {https://ui.adsabs.harvard.edu/abs/2004ApJ...616..266M} {616, 266}

\bibitem[\protect\citeauthoryear{{Maureira} et~al.,}{{Maureira}
  et~al.}{2022}]{maureira2022hotspots}
{Maureira} M.~J.,  et~al., 2022, \mn@doi [\apjl] {10.3847/2041-8213/aca53a},
  \href {https://ui.adsabs.harvard.edu/abs/2022ApJ...941L..23M} {941, L23}

\bibitem[\protect\citeauthoryear{{Mignon-Risse}, {Gonz{\'a}lez},
  {Commer{\c{c}}on}  \& {Rosdahl}}{{Mignon-Risse}
  et~al.}{2020}]{mignon-rise2020hydrid}
{Mignon-Risse} R.,  {Gonz{\'a}lez} M.,  {Commer{\c{c}}on} B.,   {Rosdahl} J.,
  2020, \mn@doi [\aap] {10.1051/0004-6361/201936605}, \href
  {https://ui.adsabs.harvard.edu/abs/2020A&A...635A..42M} {635, A42}

\bibitem[\protect\citeauthoryear{{Mouschovias} \& {Spitzer}}{{Mouschovias} \&
  {Spitzer}}{1976}]{Mouschovias1976}
{Mouschovias} T.~C.,  {Spitzer} L. J.,  1976, \mn@doi [\apj] {10.1086/154835},
  \href {https://ui.adsabs.harvard.edu/abs/1976ApJ...210..326M} {210, 326}

\bibitem[\protect\citeauthoryear{{Ohashi} et~al.,}{{Ohashi}
  et~al.}{2023}]{ohashi_tobin2023edisk}
{Ohashi} N.,  et~al., 2023, \mn@doi [\apj] {10.3847/1538-4357/acd384}, \href
  {https://ui.adsabs.harvard.edu/abs/2023ApJ...951....8O} {951, 8}

\bibitem[\protect\citeauthoryear{{Pakmor} \& {Springel}}{{Pakmor} \&
  {Springel}}{2013}]{AREPOPowell}
{Pakmor} R.,  {Springel} V.,  2013, \mn@doi [\mnras] {10.1093/mnras/stt428},
  \href {https://ui.adsabs.harvard.edu/abs/2013MNRAS.432..176P} {432, 176}

\bibitem[\protect\citeauthoryear{{Pakmor}, {Bauer}  \& {Springel}}{{Pakmor}
  et~al.}{2011}]{MHDArepo}
{Pakmor} R.,  {Bauer} A.,   {Springel} V.,  2011, \mn@doi [\mnras]
  {10.1111/j.1365-2966.2011.19591.x}, \href
  {https://ui.adsabs.harvard.edu/abs/2011MNRAS.418.1392P} {418, 1392}

\bibitem[\protect\citeauthoryear{{Pakmor}, {Springel}, {Bauer}, {Mocz},
  {Munoz}, {Ohlmann}, {Schaal}  \& {Zhu}}{{Pakmor}
  et~al.}{2016}]{pakmor2016convergene}
{Pakmor} R.,  {Springel} V.,  {Bauer} A.,  {Mocz} P.,  {Munoz} D.~J.,
  {Ohlmann} S.~T.,  {Schaal} K.,   {Zhu} C.,  2016, \mn@doi [\mnras]
  {10.1093/mnras/stv2380}, \href
  {https://ui.adsabs.harvard.edu/abs/2016MNRAS.455.1134P} {455, 1134}

\bibitem[\protect\citeauthoryear{{Pakmor} et~al.,}{{Pakmor}
  et~al.}{2023}]{mtng}
{Pakmor} R.,  et~al., 2023, \mn@doi [\mnras] {10.1093/mnras/stac3620}, \href
  {https://ui.adsabs.harvard.edu/abs/2023MNRAS.524.2539P} {524, 2539}

\bibitem[\protect\citeauthoryear{{Pakmor} et~al.,}{{Pakmor}
  et~al.}{2024}]{pakmor2024auriga}
{Pakmor} R.,  et~al., 2024, \mn@doi [\mnras] {10.1093/mnras/stae112}, \href
  {https://ui.adsabs.harvard.edu/abs/2024MNRAS.528.2308P} {528, 2308}

\bibitem[\protect\citeauthoryear{{Pandey} \& {Wardle}}{{Pandey} \&
  {Wardle}}{2008}]{Pandey2008}
{Pandey} B.~P.,  {Wardle} M.,  2008, \mn@doi [\mnras]
  {10.1111/j.1365-2966.2008.12998.x}, \href
  {https://ui.adsabs.harvard.edu/abs/2008MNRAS.385.2269P} {385, 2269}

\bibitem[\protect\citeauthoryear{{Pattle}, {Fissel}, {Tahani}, {Liu}  \&
  {Ntormousi}}{{Pattle} et~al.}{2023}]{Pattle2023}
{Pattle} K.,  {Fissel} L.,  {Tahani} M.,  {Liu} T.,   {Ntormousi} E.,  2023, in
  {Inutsuka} S.,  {Aikawa} Y.,  {Muto} T.,  {Tomida} K.,   {Tamura} M.,  eds,
  Astronomical Society of the Pacific Conference Series Vol. 534, Protostars
  and Planets VII. p.~193 (\mn@eprint {arXiv} {2203.11179}),
  \mn@doi{10.48550/arXiv.2203.11179}

\bibitem[\protect\citeauthoryear{{Pineda}, {Segura-Cox}, {Caselli},
  {Cunningham}, {Zhao}, {Schmiedeke}, {Maureira}  \& {Neri}}{{Pineda}
  et~al.}{2020}]{pineda2020streamer}
{Pineda} J.~E.,  {Segura-Cox} D.,  {Caselli} P.,  {Cunningham} N.,  {Zhao} B.,
  {Schmiedeke} A.,  {Maureira} M.~J.,   {Neri} R.,  2020, \mn@doi [Nature
  Astronomy] {10.1038/s41550-020-1150-z}, \href
  {https://ui.adsabs.harvard.edu/abs/2020NatAs...4.1158P} {4, 1158}

\bibitem[\protect\citeauthoryear{{Pineda} et~al.,}{{Pineda}
  et~al.}{2024}]{pineda2024ngc1333}
{Pineda} J.~E.,  et~al., 2024, \mn@doi [\aap] {10.1051/0004-6361/202347997},
  \href {https://ui.adsabs.harvard.edu/abs/2024A&A...686A.162P} {686, A162}

\bibitem[\protect\citeauthoryear{{Podio} et~al.,}{{Podio}
  et~al.}{2024}]{podio2024streamer}
{Podio} L.,  et~al., 2024, \mn@doi [\aap] {10.1051/0004-6361/202450742}, \href
  {https://ui.adsabs.harvard.edu/abs/2024A&A...688L..22P} {688, L22}

\bibitem[\protect\citeauthoryear{{Prole}, {Clark}, {Priestley}, {Glover}  \&
  {Regan}}{{Prole} et~al.}{2024}]{prole2024popIII}
{Prole} L.~R.,  {Clark} P.~C.,  {Priestley} F.~D.,  {Glover} S. C.~O.,
  {Regan} J.~A.,  2024, \mn@doi [The Open Journal of Astrophysics]
  {10.21105/astro.2310.10730}, \href
  {https://ui.adsabs.harvard.edu/abs/2024OJAp....7E...4P} {7, 4}

\bibitem[\protect\citeauthoryear{{Reynolds} et~al.,}{{Reynolds}
  et~al.}{2021}]{reynolds2021kinematic}
{Reynolds} N.~K.,  et~al., 2021, \mn@doi [\apjl] {10.3847/2041-8213/abcc02},
  \href {https://ui.adsabs.harvard.edu/abs/2021ApJ...907L..10R} {907, L10}

\bibitem[\protect\citeauthoryear{{Sadanari}, {Omukai}, {Sugimura}, {Matsumoto}
  \& {Tomida}}{{Sadanari} et~al.}{2023}]{sadanari2023firststars}
{Sadanari} K.~E.,  {Omukai} K.,  {Sugimura} K.,  {Matsumoto} T.,   {Tomida} K.,
   2023, \mn@doi [\mnras] {10.1093/mnras/stac3724}, \href
  {https://ui.adsabs.harvard.edu/abs/2023MNRAS.519.3076S} {519, 3076}

\bibitem[\protect\citeauthoryear{{Schneider}, {Ohlmann}, {Podsiadlowski},
  {R{\"o}pke}, {Balbus}, {Pakmor}  \& {Springel}}{{Schneider}
  et~al.}{2019}]{schneider2019mergers}
{Schneider} F. R.~N.,  {Ohlmann} S.~T.,  {Podsiadlowski} P.,  {R{\"o}pke}
  F.~K.,  {Balbus} S.~A.,  {Pakmor} R.,   {Springel} V.,  2019, \mn@doi [\nat]
  {10.1038/s41586-019-1621-5}, \href
  {https://ui.adsabs.harvard.edu/abs/2019Natur.574..211S} {574, 211}

\bibitem[\protect\citeauthoryear{{Shu}, {Adams}  \& {Lizano}}{{Shu}
  et~al.}{1987}]{Shu1987}
{Shu} F.~H.,  {Adams} F.~C.,   {Lizano} S.,  1987, \mn@doi [\araa]
  {10.1146/annurev.aa.25.090187.000323}, \href
  {https://ui.adsabs.harvard.edu/abs/1987ARA&A..25...23S} {25, 23}

\bibitem[\protect\citeauthoryear{{Springel}}{{Springel}}{2010}]{AREPO}
{Springel} V.,  2010, \mn@doi [\mnras] {10.1111/j.1365-2966.2009.15715.x},
  \href {https://ui.adsabs.harvard.edu/abs/2010MNRAS.401..791S} {401, 791}

\bibitem[\protect\citeauthoryear{{Stone}, {Tomida}, {White}  \&
  {Felker}}{{Stone} et~al.}{2020}]{stone2020athena++}
{Stone} J.~M.,  {Tomida} K.,  {White} C.~J.,   {Felker} K.~G.,  2020, \mn@doi
  [\apjs] {10.3847/1538-4365/ab929b}, \href
  {https://ui.adsabs.harvard.edu/abs/2020ApJS..249....4S} {249, 4}

\bibitem[\protect\citeauthoryear{{Tomida}, {Tomisaka}, {Matsumoto}, {Hori},
  {Okuzumi}, {Machida}  \& {Saigo}}{{Tomida}
  et~al.}{2013}]{tomida2013coreformation}
{Tomida} K.,  {Tomisaka} K.,  {Matsumoto} T.,  {Hori} Y.,  {Okuzumi} S.,
  {Machida} M.~N.,   {Saigo} K.,  2013, \mn@doi [\apj]
  {10.1088/0004-637X/763/1/6}, \href
  {https://ui.adsabs.harvard.edu/abs/2013ApJ...763....6T} {763, 6}

\bibitem[\protect\citeauthoryear{{Tomida}, {Okuzumi}  \& {Machida}}{{Tomida}
  et~al.}{2015}]{tomida2015seconddisk}
{Tomida} K.,  {Okuzumi} S.,   {Machida} M.~N.,  2015, \mn@doi [\apj]
  {10.1088/0004-637X/801/2/117}, \href
  {https://ui.adsabs.harvard.edu/abs/2015ApJ...801..117T} {801, 117}

\bibitem[\protect\citeauthoryear{{Tress} et~al.,}{{Tress}
  et~al.}{2024}]{tress2024cmz}
{Tress} R.~G.,  et~al., 2024, \mn@doi [arXiv e-prints]
  {10.48550/arXiv.2403.13048}, \href
  {https://ui.adsabs.harvard.edu/abs/2024arXiv240313048T} {p. arXiv:2403.13048}

\bibitem[\protect\citeauthoryear{{Tsukamoto}}{{Tsukamoto}}{2024}]{tsukamoto2024grainsII}
{Tsukamoto} Y.,  2024, \mn@doi [arXiv e-prints] {10.48550/arXiv.2404.13843},
  \href {https://ui.adsabs.harvard.edu/abs/2024arXiv240413843T} {p.
  arXiv:2404.13843}

\bibitem[\protect\citeauthoryear{{Tsukamoto}, {Takahashi}, {Machida}  \&
  {Inutsuka}}{{Tsukamoto} et~al.}{2015a}]{tsukamoto2015radiative}
{Tsukamoto} Y.,  {Takahashi} S.~Z.,  {Machida} M.~N.,   {Inutsuka} S.,  2015a,
  \mn@doi [\mnras] {10.1093/mnras/stu2160}, \href
  {https://ui.adsabs.harvard.edu/abs/2015MNRAS.446.1175T} {446, 1175}

\bibitem[\protect\citeauthoryear{{Tsukamoto}, {Iwasaki}, {Okuzumi}, {Machida}
  \& {Inutsuka}}{{Tsukamoto} et~al.}{2015b}]{tsukamoto2015q2b}
{Tsukamoto} Y.,  {Iwasaki} K.,  {Okuzumi} S.,  {Machida} M.~N.,   {Inutsuka}
  S.,  2015b, \mn@doi [\mnras] {10.1093/mnras/stv1290}, \href
  {https://ui.adsabs.harvard.edu/abs/2015MNRAS.452..278T} {452, 278}

\bibitem[\protect\citeauthoryear{{Tsukamoto}, {Iwasaki}, {Okuzumi}, {Machida}
  \& {Inutsuka}}{{Tsukamoto} et~al.}{2015c}]{tsukamoto2015q2a}
{Tsukamoto} Y.,  {Iwasaki} K.,  {Okuzumi} S.,  {Machida} M.~N.,   {Inutsuka}
  S.,  2015c, \mn@doi [\apjl] {10.1088/2041-8205/810/2/L26}, \href
  {https://ui.adsabs.harvard.edu/abs/2015ApJ...810L..26T} {810, L26}

\bibitem[\protect\citeauthoryear{{Tsukamoto}, {Okuzumi}, {Iwasaki}, {Machida}
  \& {Inutsuka}}{{Tsukamoto} et~al.}{2017}]{tsukamoto2017hall}
{Tsukamoto} Y.,  {Okuzumi} S.,  {Iwasaki} K.,  {Machida} M.~N.,   {Inutsuka}
  S.-i.,  2017, \mn@doi [\pasj] {10.1093/pasj/psx113}, \href
  {https://ui.adsabs.harvard.edu/abs/2017PASJ...69...95T} {69, 95}

\bibitem[\protect\citeauthoryear{{Tsukamoto}, {Okuzumi}, {Iwasaki}, {Machida}
  \& {Inutsuka}}{{Tsukamoto} et~al.}{2018}]{tsukamoto2018misalignment}
{Tsukamoto} Y.,  {Okuzumi} S.,  {Iwasaki} K.,  {Machida} M.~N.,   {Inutsuka}
  S.,  2018, \mn@doi [\apj] {10.3847/1538-4357/aae4dc}, \href
  {https://ui.adsabs.harvard.edu/abs/2018ApJ...868...22T} {868, 22}

\bibitem[\protect\citeauthoryear{{Tsukamoto}, {Machida}, {Susa}, {Nomura}  \&
  {Inutsuka}}{{Tsukamoto} et~al.}{2020}]{tsukamoto2020dustmodel}
{Tsukamoto} Y.,  {Machida} M.~N.,  {Susa} H.,  {Nomura} H.,   {Inutsuka} S.,
  2020, \mn@doi [\apj] {10.3847/1538-4357/ab93d0}, \href
  {https://ui.adsabs.harvard.edu/abs/2020ApJ...896..158T} {896, 158}

\bibitem[\protect\citeauthoryear{{Tsukamoto}, {Machida}  \&
  {Inutsuka}}{{Tsukamoto} et~al.}{2023a}]{tsukamoto2023grainsI}
{Tsukamoto} Y.,  {Machida} M.~N.,   {Inutsuka} S.-i.,  2023a, \mn@doi [\pasj]
  {10.1093/pasj/psad040}, \href
  {https://ui.adsabs.harvard.edu/abs/2023PASJ...75..835T} {75, 835}

\bibitem[\protect\citeauthoryear{{Tsukamoto} et~al.,}{{Tsukamoto}
  et~al.}{2023b}]{tsukamoto2023ppvii}
{Tsukamoto} Y.,  et~al., 2023b, in {Inutsuka} S.,  {Aikawa} Y.,  {Muto} T.,
  {Tomida} K.,   {Tamura} M.,  eds,  Astronomical Society of the Pacific
  Conference Series Vol. 534, Protostars and Planets VII. p.~317 (\mn@eprint
  {arXiv} {2209.13765}), \mn@doi{10.48550/arXiv.2209.13765}

\bibitem[\protect\citeauthoryear{{Tu}, {Li}, {Zhu}  \& {Hsu}}{{Tu}
  et~al.}{2024}]{tu2024drod}
{Tu} Y.,  {Li} Z.-Y.,  {Zhu} Z.,   {Hsu} C.-Y.,  2024, \mn@doi [\mnras]
  {10.1093/mnras/stae1639}, \href
  {https://ui.adsabs.harvard.edu/abs/2024MNRAS.532.3135T} {532, 3135}

\bibitem[\protect\citeauthoryear{{Valdivia-Mena} et~al.,}{{Valdivia-Mena}
  et~al.}{2022}]{valdivia-mena2022streamer}
{Valdivia-Mena} M.~T.,  et~al., 2022, \mn@doi [\aap]
  {10.1051/0004-6361/202243310}, \href
  {https://ui.adsabs.harvard.edu/abs/2022A&A...667A..12V} {667, A12}

\bibitem[\protect\citeauthoryear{{Vaytet}, {Commer{\c{c}}on}, {Masson},
  {Gonz{\'a}lez}  \& {Chabrier}}{{Vaytet}
  et~al.}{2018}]{vaytet2018protostellar}
{Vaytet} N.,  {Commer{\c{c}}on} B.,  {Masson} J.,  {Gonz{\'a}lez} M.,
  {Chabrier} G.,  2018, \mn@doi [\aap] {10.1051/0004-6361/201732075}, \href
  {https://ui.adsabs.harvard.edu/abs/2018A&A...615A...5V} {615, A5}

\bibitem[\protect\citeauthoryear{{Vogelsberger} et~al.,}{{Vogelsberger}
  et~al.}{2014}]{Vogelsberger2014}
{Vogelsberger} M.,  et~al., 2014, \mn@doi [\mnras] {10.1093/mnras/stu1536},
  \href {https://ui.adsabs.harvard.edu/abs/2014MNRAS.444.1518V} {444, 1518}

\bibitem[\protect\citeauthoryear{{Weinberger}, {Springel}  \&
  {Pakmor}}{{Weinberger} et~al.}{2020}]{AREPOpublic}
{Weinberger} R.,  {Springel} V.,   {Pakmor} R.,  2020, \mn@doi [\apjs]
  {10.3847/1538-4365/ab908c}, \href
  {https://ui.adsabs.harvard.edu/abs/2020ApJS..248...32W} {248, 32}

\bibitem[\protect\citeauthoryear{{Whitehouse} \& {Bate}}{{Whitehouse} \&
  {Bate}}{2006}]{whitehouse_bate2006collapse}
{Whitehouse} S.~C.,  {Bate} M.~R.,  2006, \mn@doi [\mnras]
  {10.1111/j.1365-2966.2005.09950.x}, \href
  {https://ui.adsabs.harvard.edu/abs/2006MNRAS.367...32W} {367, 32}

\bibitem[\protect\citeauthoryear{{Wurster}}{{Wurster}}{2016}]{wurster2016nicil}
{Wurster} J.,  2016, \mn@doi [\pasa] {10.1017/pasa.2016.34}, \href
  {https://ui.adsabs.harvard.edu/abs/2016PASA...33...41W} {33, e041}

\bibitem[\protect\citeauthoryear{{Wurster}}{{Wurster}}{2021a}]{wurster2021doweneed}
{Wurster} J.,  2021a, \mn@doi [\mnras] {10.1093/mnras/staa3943}, \href
  {https://ui.adsabs.harvard.edu/abs/2021MNRAS.501.5873W} {501, 5873}

\bibitem[\protect\citeauthoryear{{Wurster}}{{Wurster}}{2021b}]{wurster2021nicil2.0}
{Wurster} J.,  2021b, \mn@doi [\mnras] {10.1093/mnras/staa3943}, \href
  {https://ui.adsabs.harvard.edu/abs/2021MNRAS.501.5873W} {501, 5873}

\bibitem[\protect\citeauthoryear{{Wurster} \& {Lewis}}{{Wurster} \&
  {Lewis}}{2020}]{wurster2020turbI}
{Wurster} J.,  {Lewis} B.~T.,  2020, \mn@doi [\mnras] {10.1093/mnras/staa1339},
  \href {https://ui.adsabs.harvard.edu/abs/2020MNRAS.495.3795W} {495, 3795}

\bibitem[\protect\citeauthoryear{{Wurster}, {Price}  \& {Ayliffe}}{{Wurster}
  et~al.}{2014}]{wurster2014ambipolar}
{Wurster} J.,  {Price} D.,   {Ayliffe} B.,  2014, \mn@doi [\mnras]
  {10.1093/mnras/stu1524}, \href
  {https://ui.adsabs.harvard.edu/abs/2014MNRAS.444.1104W} {444, 1104}

\bibitem[\protect\citeauthoryear{{Wurster}, {Price}  \& {Bate}}{{Wurster}
  et~al.}{2016}]{wurster2016cannonidealmhd}
{Wurster} J.,  {Price} D.~J.,   {Bate} M.~R.,  2016, \mn@doi [\mnras]
  {10.1093/mnras/stw013}, \href
  {https://ui.adsabs.harvard.edu/abs/2016MNRAS.457.1037W} {457, 1037}

\bibitem[\protect\citeauthoryear{{Wurster}, {Bate}  \& {Price}}{{Wurster}
  et~al.}{2018a}]{wurster2018collapse}
{Wurster} J.,  {Bate} M.~R.,   {Price} D.~J.,  2018a, \mn@doi [\mnras]
  {10.1093/mnras/stx3339}, \href
  {https://ui.adsabs.harvard.edu/abs/2018MNRAS.475.1859W} {475, 1859}

\bibitem[\protect\citeauthoryear{{Wurster}, {Bate}  \& {Price}}{{Wurster}
  et~al.}{2018b}]{wurster2018ionizationrates}
{Wurster} J.,  {Bate} M.~R.,   {Price} D.~J.,  2018b, \mn@doi [\mnras]
  {10.1093/mnras/sty392}, \href
  {https://ui.adsabs.harvard.edu/abs/2018MNRAS.476.2063W} {476, 2063}

\bibitem[\protect\citeauthoryear{{Wurster}, {Bate}  \& {Price}}{{Wurster}
  et~al.}{2019}]{wurster2019catastrophe}
{Wurster} J.,  {Bate} M.~R.,   {Price} D.~J.,  2019, \mn@doi [\mnras]
  {10.1093/mnras/stz2215}, \href
  {https://ui.adsabs.harvard.edu/abs/2019MNRAS.489.1719W} {489, 1719}

\bibitem[\protect\citeauthoryear{{Wurster}, {Bate}  \& {Bonnell}}{{Wurster}
  et~al.}{2021}]{wurster2021nonidealimpactsingle}
{Wurster} J.,  {Bate} M.~R.,   {Bonnell} I.~A.,  2021, \mn@doi [\mnras]
  {10.1093/mnras/stab2296}, \href
  {https://ui.adsabs.harvard.edu/abs/2021MNRAS.507.2354W} {507, 2354}

\bibitem[\protect\citeauthoryear{Wurster, Bate, Price  \& Bonnell}{Wurster
  et~al.}{2022}]{wurster2022resolution}
Wurster J.,  Bate M.~R.,  Price D.~J.,   Bonnell I.~A.,  2022, \mn@doi [Monthly
  Notices of the Royal Astronomical Society] {10.1093/mnras/stac123}, 511, 746

\bibitem[\protect\citeauthoryear{{Xu} \& {Kunz}}{{Xu} \&
  {Kunz}}{2021a}]{xu2021formationI}
{Xu} W.,  {Kunz} M.~W.,  2021a, \mn@doi [\mnras] {10.1093/mnras/stab314}, \href
  {https://ui.adsabs.harvard.edu/abs/2021MNRAS.502.4911X} {502, 4911}

\bibitem[\protect\citeauthoryear{{Xu} \& {Kunz}}{{Xu} \&
  {Kunz}}{2021b}]{xu2021formationII}
{Xu} W.,  {Kunz} M.~W.,  2021b, \mn@doi [\mnras] {10.1093/mnras/stab2715},
  \href {https://ui.adsabs.harvard.edu/abs/2021MNRAS.508.2142X} {508, 2142}

\bibitem[\protect\citeauthoryear{{Yen} \& {Lee}}{{Yen} \&
  {Lee}}{2024}]{yen_lee2024braking}
{Yen} H.-W.,  {Lee} Y.-N.,  2024, arXiv e-prints, \href
  {https://ui.adsabs.harvard.edu/abs/2024arXiv240812101Y} {p. arXiv:2408.12101}

\bibitem[\protect\citeauthoryear{{Zhao}, {Caselli}, {Li}, {Krasnopolsky},
  {Shang}  \& {Nakamura}}{{Zhao} et~al.}{2016}]{Zhao2016}
{Zhao} B.,  {Caselli} P.,  {Li} Z.-Y.,  {Krasnopolsky} R.,  {Shang} H.,
  {Nakamura} F.,  2016, \mn@doi [\mnras] {10.1093/mnras/stw1124}, \href
  {https://ui.adsabs.harvard.edu/abs/2016MNRAS.460.2050Z} {460, 2050}

\bibitem[\protect\citeauthoryear{{Zhao} et~al.,}{{Zhao}
  et~al.}{2020a}]{zhao2020review}
{Zhao} B.,  et~al., 2020a, \mn@doi [\ssr] {10.1007/s11214-020-00664-z}, \href
  {https://ui.adsabs.harvard.edu/abs/2020SSRv..216...43Z} {216, 43}

\bibitem[\protect\citeauthoryear{{Zhao}, {Caselli}, {Li}, {Krasnopolsky},
  {Shang}  \& {Lam}}{{Zhao} et~al.}{2020b}]{zhao2020hall}
{Zhao} B.,  {Caselli} P.,  {Li} Z.-Y.,  {Krasnopolsky} R.,  {Shang} H.,   {Lam}
  K.~H.,  2020b, \mn@doi [\mnras] {10.1093/mnras/staa041}, \href
  {https://ui.adsabs.harvard.edu/abs/2020MNRAS.492.3375Z} {492, 3375}

\bibitem[\protect\citeauthoryear{{Zhao}, {Caselli}, {Li}, {Krasnopolsky},
  {Shang}  \& {Lam}}{{Zhao} et~al.}{2021}]{zhao2021interplay}
{Zhao} B.,  {Caselli} P.,  {Li} Z.-Y.,  {Krasnopolsky} R.,  {Shang} H.,   {Lam}
  K.~H.,  2021, \mn@doi [\mnras] {10.1093/mnras/stab1295}, \href
  {https://ui.adsabs.harvard.edu/abs/2021MNRAS.505.5142Z} {505, 5142}

\bibitem[\protect\citeauthoryear{{Zier} \& {Springel}}{{Zier} \&
  {Springel}}{2022a}]{AREPOcolddisks}
{Zier} O.,  {Springel} V.,  2022a, \mn@doi [\mnras] {10.1093/mnras/stac1783},
  \href {https://ui.adsabs.harvard.edu/abs/2022MNRAS.515..525Z} {515, 525}

\bibitem[\protect\citeauthoryear{{Zier} \& {Springel}}{{Zier} \&
  {Springel}}{2022b}]{AREPOmri}
{Zier} O.,  {Springel} V.,  2022b, \mn@doi [\mnras] {10.1093/mnras/stac2831},
  \href {https://ui.adsabs.harvard.edu/abs/2022MNRAS.517.2639Z} {517, 2639}

\bibitem[\protect\citeauthoryear{{Zier} \& {Springel}}{{Zier} \&
  {Springel}}{2023}]{arepoFragmentation}
{Zier} O.,  {Springel} V.,  2023, \mn@doi [\mnras] {10.1093/mnras/stad319},
  \href {https://ui.adsabs.harvard.edu/abs/2023MNRAS.520.3097Z} {520, 3097}

\bibitem[\protect\citeauthoryear{{Zier}, {Springel}  \& {Mayer}}{{Zier}
  et~al.}{2024a}]{AREPODiffusion}
{Zier} O.,  {Springel} V.,   {Mayer} A.~C.,  2024a, \mn@doi [\mnras]
  {10.1093/mnras/stad3200}, \href
  {https://ui.adsabs.harvard.edu/abs/2024MNRAS.527.1563Z} {527, 1563}

\bibitem[\protect\citeauthoryear{{Zier}, {Mayer}  \& {Springel}}{{Zier}
  et~al.}{2024b}]{AREPOHall}
{Zier} O.,  {Mayer} A.~C.,   {Springel} V.,  2024b, \mn@doi [\mnras]
  {10.1093/mnras/stad3769}, \href
  {https://ui.adsabs.harvard.edu/abs/2024MNRAS.527.8355Z} {527, 8355}

\makeatother
\end{thebibliography}

\appendix

\section{Rotational structure close to stellar core formation}
\label{app:rotationalStructure}

\begin{figure*}
    \centering
    \includegraphics[width=1\linewidth]{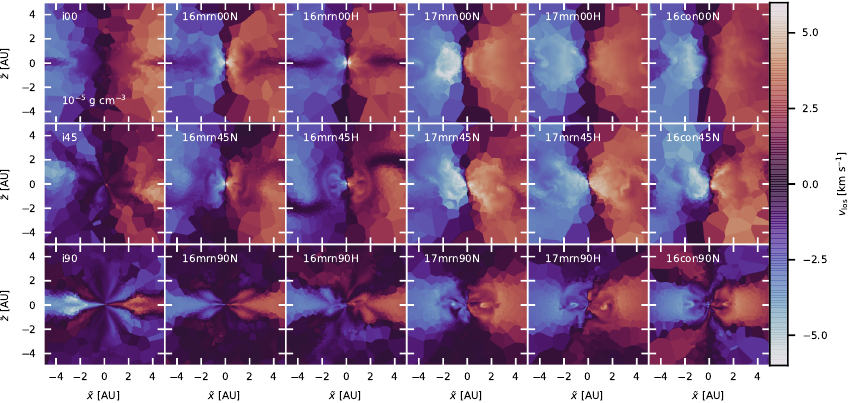}
    \caption{Edge-on line-of-sight velocity close to the time of second core formation for all but the 16conH runs. Small rotating regions have formed in 16mrn00 and 16mrn45.}
    \label{fig:los_late}
\end{figure*}

\begin{figure*}
    \centering
    \includegraphics[width=1\linewidth]{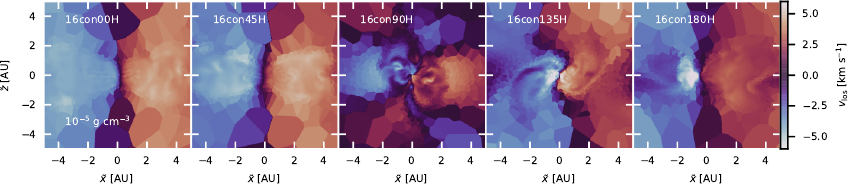}
    \caption{Edge-on line-of-sight velocity close to the time of second core formation for the 16conH runs. The twist in 16con90H propagates down to the scale of the second core.}
    \label{fig:los_late_hall}
\end{figure*}

Figures~\ref{fig:los_late} and \ref{fig:los_late_hall} focus on the velocity structure at later times, and they are turned such that the $\tilde{z}$-axis is aligned with the angular momentum vector of the first core, and the $\tilde{x}$-axis is approximately perpendicular to the pseudo-disk. At this point, even some of the models which had not formed rotationally supported disks in the first core phase display a small but fast-rotating central region, most clearly 16mrn00N/H and 16mrn45N/H. In the ideal MHD models and those with $90^\circ$-inclination, on the other hand, there is essentially no centrifugal support to stop the collapse, and the central region has a less well-ordered rotation profile.

\section{Comparison of Poynting and kinetic flux}
\label{app:poyntingVsKineticFlux}

\begin{figure*}
    \centering
    \includegraphics[width=1\linewidth]{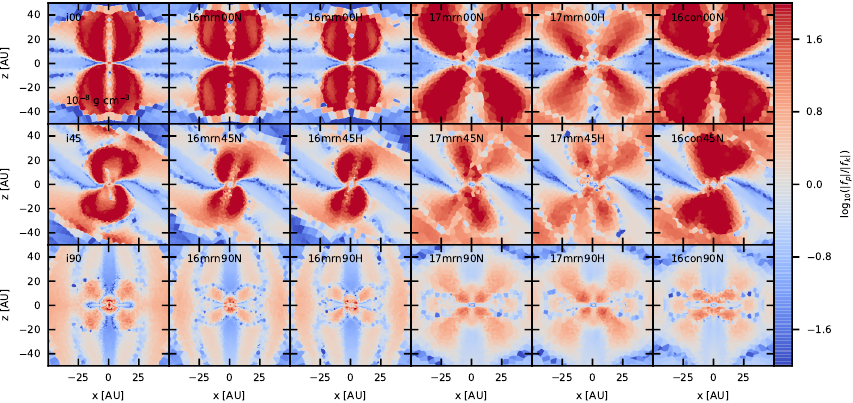}
    \caption{$xz$-slices of the ratios of Poynting- to kinetic flux at the time the simulation reaches $\rho_{\rm max} = 10^{-8} \, {\rm g} \, {\rm cm}^{-3}$ for the first time. Outflow regions are heavily magnetically dominated.}
    \label{fig:poynting_all}
\end{figure*}

\begin{figure*}
    \centering
    \includegraphics[width=1\linewidth]{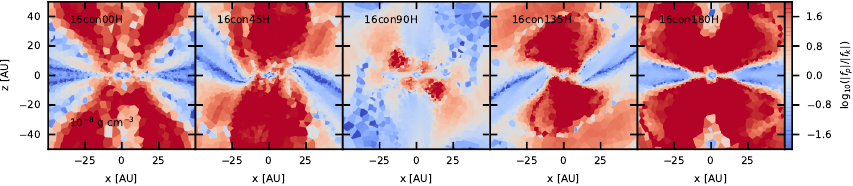}
    \caption{$xz$-slices of the ratios of Poynting- to kinetic flux at the time the simulation reaches $\rho_{\rm max} = 10^{-8} \, {\rm g} \, {\rm cm}^{-3}$ for the first time.}
    \label{fig:poynting_hall}
\end{figure*}

Similar to, e.g., \cite{bate2014collapse} and \cite{wurster2021nonidealimpactsingle}, in Figures~\ref{fig:poynting_all} and ~\ref{fig:poynting_hall} we compare the radial components of the Poynting flux ${\bm f}_{\rm P} = {\bm B}\times (\mathbf{v} \times {\bm B})$ and the kinetic flux ${\bm f}_{\rm k} = \frac{1}{2} \rho \vert \mathbf{v} \vert^2 \mathbf{v}$ to ascertain whether the outflows in our models are magnetically dominated. As can be seen in the Figures, the Poynting flux dominates in all outflow regions, but this is not exclusive to the outflows. In particular, even simulations with no outflows also have large regions where the (radial) Poynting flux dominates, and these regions appear to nicely map out the boundary of the pseudo-disk with the surrounding envelope in all simulations. The actual disk plane is then again dominated by the kinetic flux.

\section{Modifications to the non-ideal MHD scheme}
\label{app:modificationsNonIdealMHD}

We have slightly modified the non-ideal MHD modules with respect to the original implementation. The modification is related to the fallback strategy in the case of a potentially unstable least-square-fit (LSF) over all cells touching the interface, as described in section~2.3.1 of \citet{AREPODiffusion}. This is because setting the components of the magnetic field gradient parallel to the interface to zero systematically introduces divergence errors, as at the interface one always has $\nabla \cdot {\bm B} = \frac{\partial B_{\rm n}}{\partial x_{\rm n}}$, where $B_{\rm n}$ and $x_{\rm n}$ refer to the components normal to the interface. This leads to instabilities which influence the overall behavior of the simulation in regions where the non-ideal MHD fluxes dominate. For the normal component, the finite-differences gradient estimate used in the previous version of the fallback is in the best case scenario correct to second order if the mesh-generating points are at the same positions as the centers of masses on both sides of the interface. 

However, in this best-case scenario for the normal component, the equivalent estimate for the parallel components (where the separation vector is projected onto the vectors spanning the interface) is necessarily equal to zero. Since {\small AREPO} regularizes the mesh by slowly moving the mesh-generating points towards the center of mass of a cell if their separation is too large, the parallel component of the gradient in this finite-differences fallback will therefore be systematically underestimated in comparison to the normal component. Due to the structure of the flux-terms for ambipolar diffusion and the Hall-effect, this does not in general correspond to an underestimation of the fluxes (i.e. the fallback would be acting as a flux-limiter), but can rather be an overestimate and changes the nature of the terms. Therefore, we have decided to not utilize this fallback strategy and instead we improve the quality of the mesh (and therefore the quality of the LSF fit) by using somewhat more aggressive mesh-regularization. As the tests performed in \cite{AREPODiffusion} and \cite{AREPOHall} demonstrate, the convergence of our scheme is not noticeably worsened by the use of the fallback strategy, but this relies on the fact that the constantly moving mesh will keep interfaces from having many inaccurate flux calculations repeatedly due to a problematic surrounding geometry. 

The speed of mesh-regularization motions is normally limited to the largest hydrodynamical speed of the cell (i.e. either its soundspeed or velocity), but the timescales associated with the non-ideal MHD fluxes (see equation \eqref{eq: timestep_constraint}) can be orders of magnitude smaller than the Alfv{\' e}n timescale, which leads to a mesh that is essentially unchanged over many non-ideal MHD calculations. Therefore, geometries leading to inaccuracies in the LSF can here persist and cause instabilities. We thus now allow mesh-motions to be as fast as the timescale associated with the non-ideal MHD. We have also tested further increasing the quality of the mesh by enforcing rounder cells (reducing the \textit{CellMaxAngleFactor} parameter of {\small AREPO}), which we find to significantly reduce the extrapolation in the LSF fit; however, this leads to larger deviations from Lagrangian behaviour and thereby to more diffusive behavior. We therefore decided not to use it in this work, as the simulations performed are stable without it. We have also found that other methods that tend to increase diffusion can provide stability in situations more extreme than treated here, such as using the Dedner divergence cleaning scheme that was implemeted originally in {\small AREPO}, or adding artificial Ohmic diffusion, but they come at a similar cost to accuracy in terms of added numerical resistivity.

\bsp	
\label{lastpage}
\end{document}